\documentclass{article}
\usepackage[utf8]{inputenc}

\usepackage{graphics}
\usepackage{graphicx}
\usepackage{mathtools}

\usepackage{geometry}

\usepackage{makecell}
\usepackage{multirow} 
\usepackage[table]{xcolor} 

\usepackage{xcolor}
\usepackage{amsmath}
\usepackage[allcolors=false, hidelinks]{hyperref}
\begin{document}

\title{
Morphology Formation Pathways in Solution-Processed Perovskite Thin Films

}
\maketitle

M. Majewski, O.J.J. Ronsin, J. Harting*\\
Helmholtz Institute Erlangen-Nürnberg for Renewable Energy (HIERN), Forschungszentrum Jülich GmbH\\
Email Address: j.harting@fz-juelich.de

J. Harting\\
Department of Chemical and Biological Engineering and Department of Physics, Friedrich-Alexander-Universität Erlangen-Nürnberg, Fürther Straße 248, 90429 Nürnberg, Germany

\begin{abstract}
The active layer in a perovskite solar cell is usually composed of a polycrystalline thin film. Fabrication of this layer by solution processing is a promising candidate for up-scaling to the mass market. 
However, the evolution of an evaporating and simultaneously crystallizing thin film is not yet fully understood. To contribute to the understanding of the formation of thin films, we develop a geometrical model that deals with the effect of the
interplay between solvent evaporation and crystal growth on the dry film morphology. The
possible film formation mechanisms are investigated, depending on the processing conditions. We find eleven
formation pathways leading to four distinct morphologies. It is shown how these
formation pathways can be utilized  
by adapting the process parameters to the material properties. Pinhole-free and flat films can be
fabricated if the evaporation rate is high in comparison to the crystal growth
rate. Alternatively, providing a high crystal number density on the substrate
can lead to the desired film morphology at low drying rates. The generality of the model makes it applicable to any evaporating and simultaneously crystallizing thin film.
\end{abstract}

\section{Introduction}

Perovskite (PSK) solar cells have the potential to become the next generation of solar cells, reaching the mass market~\cite{noauthor_best_nodate-1,zhu_toward_2024}. Their advantages include high power conversion efficiency, outstanding optoelectrical properties, and a low-cost, low-energy fabrication route~\cite{schmidt-mende_roadmap_2021}. However, reliable deposition of high-quality films remains a challenge~\cite{parida_recent_2022,dai_meniscus_2019,dunlap-shohl_synthetic_2019}. Addressing this issue necessitates process design rules, based on physical understanding, for fully closed, large-area, smooth films. This is not straightforward as PSK film growth is complex and involves sophisticated solution chemistry~\cite{shargaieva_temperature-dependent_2021,jiao_solvent_2023}.

Different processing routes are possible, where deposition from solution is a promising candidate with potential for scale-up~\cite{yang_upscaling_2021,wang_solution-processable_2019,liu_controlling_2022}. In this approach, the physical processes that occur within a drying and simultaneously crystallizing film need to be understood. In principle, a mixture of precursor materials is dissolved, and a wet film is deposited on a substrate. Upon solvent evaporation, the precursor materials crystallize into perovskite. The perovskite crystals may form either directly, or the precursor materials may form crystalline intermediates with the solvent, from which the perovskite itself crystallizes in a second step~\cite{stone_transformation_2018,valencia_optical_2021}. In this process, the amount of the intermediate phase primarily depends on the solvent~\cite{shargaieva_temperature-dependent_2021}. 
For weakly binding solvents, such as $\gamma$-Butyrolacton (GBL) or 2-Methoxyethanol (2-ME), the transformation into perovskite was shown to be direct~\cite{dai_meniscus_2019,witt_orientation_2023}. In contrast, for strongly binding solvents, the intermediate phase quickly crystallizes and determines the final morphology~\cite{valencia_optical_2021,yan_hybrid_2015,li_phase_2018}. The specific formation pathway that results from a combination of solvents and a perovskite material system is an ongoing topic of research~\cite{unlu_toward_2025, gallant_green_2024}.

Experimentally, it was observed that a full substrate coverage and a low surface roughness are crucial for high-efficiency devices~\cite{li_highly_2023,qiu_pinhole-free_2016}. In the case of the metal-halide perovskite materials used in printed photovoltaics, it was observed that drying has to be fast in order to obtain high-quality films, which might appear counter-intuitive. This highlights how crucial it is to understand how processing conditions, such as the initial precursor concentration, temperature, and evaporation rate, influence the film formation mechanisms and the final morphology~\cite{ternes_drying_2022,guo_high_2022}. However, the morphology formation pathway is difficult to determine experimentally~\cite{telschow_elucidating_2023}. It can be estimated using analytical models~\cite{he_meniscus-assisted_2017}, such as the Lamer model~\cite{ding_material_2017,jiao_solvent_2023}, or through numerical simulations. Phase-field simulations allowed for describing the multi-component demixing of blends in drying films for organic photovoltaic systems~\cite{negi_simulating_2018,zhao_vertical_2016,wodo_modeling_2012}. Similar methods were also applied to investigate the process-morphology relationships in meniscus-guided doctor blading for crystallizing films in the evaporative regime, and for immiscible blends in the Landau-Levich regime~\cite{de_bruijn_periodic_2025,michels_predictive_2021,de_bruijn_structuring_2024}. Furthermore, evaporating crystallizing thin films were studied in the context of drug release~\cite{kim_modeling_2009,saylor_predicting_2011,saylor_diffuse_2016}. Recently, a theoretical description of the interplay between the nucleation and the evaporation rate of a thin film was proposed~\cite{de_bruijn_transient_2024}, while others developed a model for predicting the supersaturation rate, which is expected to drive perovskite crystallization in drying films~\cite{ternes_modeling_nodate}. 
Nevertheless, the evolution of a doctor-bladed film in the Landau-Levich regime is not fully understood yet. Especially the necessity to dry fast in order to obtain a functional active layer in a perovskite solar cell, was unexpected~\cite{dunlap-shohl_synthetic_2019}.

To this end, our group developed a phase-field simulation framework~\cite{ronsin_phase-field_2022} that allows us to investigate the morphology formation of both organic~\cite{ronsin_formation_2022} and perovskite~\cite{majewski_simulation_2025} photo-active layers during wet film drying. We were able to validate the simulations by extensive comparison with experiments. Additionally, we found that the ratio of solvent evaporation rate to crystallization rate is the determining factor for the morphology formation. It has to be high in order to achieve a high-quality, smooth perovskite active layer~\cite{majewski_simulation_2025}. Furthermore, we showed that favoring perovskite growth from the substrate can also support the film quality. Thereby, it might become favorable to process at lower drying rates~\cite{qiu_over_2025}. In parallel to the simulations, analytical models need to be developed in order to obtain a broad and conclusive insight into the possible formation pathways.

Here, we develop a model that describes the growth of crystals located on the substrate in drying films. It allows for the identification of important events during film formation. The order in which they occur defines the formation pathway of the film and determines the final morphology. Although being very simplistic, the model allows us to establish generic guidelines for the real-world problem of depositing a smooth perovskite layer on a large area. Since no material-specific assumptions are made, these guidelines also apply to any other systems with similar properties.

In Section 2, we describe the assumptions underlying the model. Then, we discuss all the possible film formation pathways in the model in Section 3. In section 4, we explore the role of the drying rate and the crystal number density on the formation pathways and morphologies, and we connect our findings to the deposition of real films. In Section 5, we then examine the effects of the initial volume fraction of the wet film and of the initial crystal size. The paper concludes with a summary and an outlook.

\section{Model}\label{sec:model}

A precursor solution is deposited in a single step using a suitable method, such as doctor blading, spin coating, or slot-die coating. Immediately after deposition, the precursor materials are dissolved, and the film is a homogeneous liquid mixture. We consider the following two common experimental scenarios: {First, the precursor solution is deposited on a substrate that is patterned with seeds. There is total control over the size and the place of all the seeds. Second, the precursor solution is deposited on a blank substrate. Heterogeneous nucleation at the substrate is ensured to be dominant compared to homogeneous nucleation within the film. In this case, the initial crystal sizes and the patterning density are statistically distributed and defined by the material properties. In this work, we represent heterogeneous nucleation through the average nucleation density and the critical radius. These properties, together with the processing conditions, define the final film morphology.} 
In both cases, as the wet film subsequently dries, the wet film height decreases and the perovskite crystals grow from the substrate. 
The evaporation rate of the solvent can be controlled (e.g., by gas quenching). The growth rate of the crystals is material-specific and therefore harder to modify. The initial volume and composition can be tuned up to the saturation threshold of the solute in the solvent.

\begin{figure}[h]
\centering
  \includegraphics[width = 0.49\textwidth]{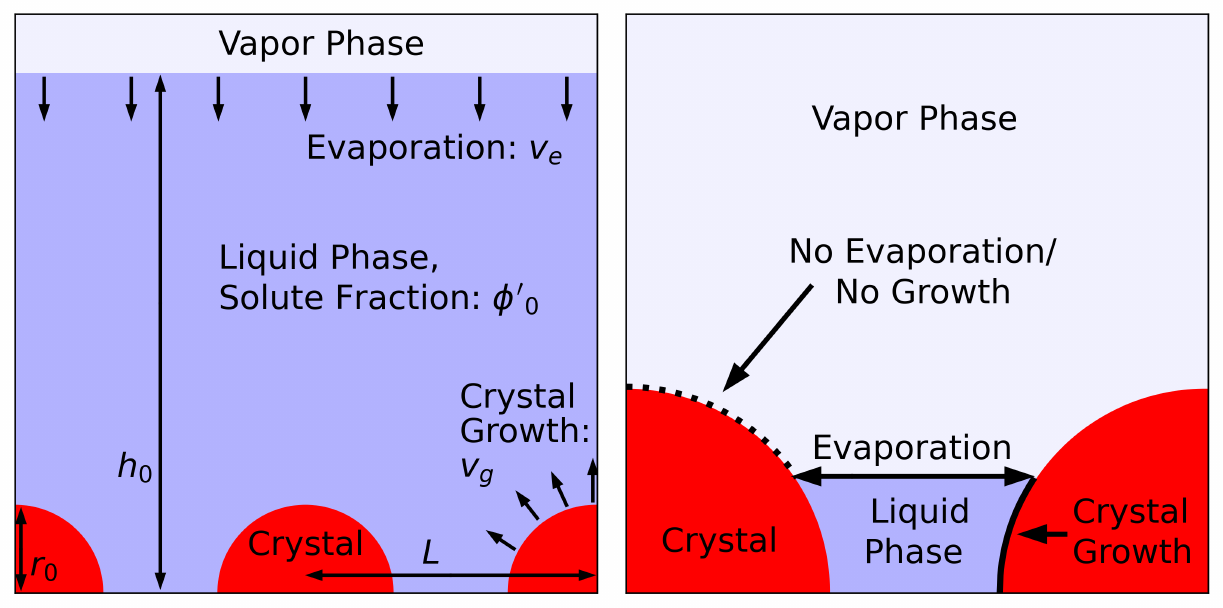}
  \caption{Left: Sketch of the initial state of the considered film, including model parameters. The vapor phase is shown in light blue, the liquid phase in dark blue and the crystalline phase in red. Right: Sketch of an intermediate state (zoom in) highlighting where crystal growth and solvent evaporation can take place.}
  \label{fig:setup}
\end{figure}

Our model is designed to mimic this situation in a simplified manner. We consider the 2D cross-section of a drying film with half-spherical crystals initially located on an inert substrate and a vapor layer above it, as depicted in \autoref{fig:setup}. We investigate periodic arrangements of the crystals in the horizontal direction, with a unit cell consisting of two crystals of equal size and with equal distances between them. The evolution of both the crystalline and the liquid phase can then be modelled. The liquid phase is composed of one solvent and one solute. Note that the complex chemistry involved in perovskite crystallization is not taken into account, and the solute represents the material that eventually crystallizes. Only the solvent evaporates, while both the solute and crystalline phase cannot. The composition of the liquid phase is fixed initially but evolves with time. The liquid phase is assumed to always remain homogeneous, which implies that material transport is not diffusion-limited. The liquid-vapor interface velocity (hereafter referred to as the evaporation rate) arising due to evaporation is modeled as constant for simplicity, until evaporation globally stops when there is no solvent left. The crystalline phase is composed of pure solute and grows at a constant interface velocity (hereafter referred to as the crystal growth rate). Crystal growth consumes solute from the liquid phase, and hence globally comes to a halt when the liquid phase consists of pure solvent only. In addition, crystal growth and evaporation stop locally wherever the crystal surface is exposed to air, as shown in \autoref{fig:setup}~(right). Surface tensions and/or wetting effects are not taken into account, so that the surface of the liquid phase is assumed to always remain flat. It is assumed that further nucleation during film drying is successfully avoided, so that crystallization proceeds only through growth of the initial seeds. The proposed picture is simple enough to be analytically tractable. Complexifying the model, for example by including diffusion-limited mass transport, and non-constant evaporation and growth rates, would quickly become analytically very challenging. However, phase-field simulations (see SI, section 8) show that such refinements would not qualitatively change the findings presented in this work.

The model parameters are as follows: $r_{0}$ is the radius of the initially placed (half-spherical) crystals, $h_{0}$ is the initial height of the liquid film, and $\phi_{0}$ is the initial fraction of solute in the condensed film. The distance between the centers of the crystals is given by $L$, while $v_e$ and $v_g$ describe the solvent evaporation rate and the crystal growth rate, respectively. These parameters are visualized in \autoref{fig:setup}~(left). In order to reduce the dimensionality of the parameter space, in the following we discuss the behavior of the drying film depending on the four independent parameters $v_e/v_g$, $L/h_0$, $r_0/h_0$ and $\phi_0$.

The initial volume fraction $\phi_0$ is defined as the volume fraction in the condensed film (which is the liquid and the solid phase). This may seem counter intuitive, but it has the advantage of simplified equations. In the case of dominant heterogeneous nucleation at the substrate, it is the volume fraction of the amorphous phase at the moment when the system is still liquid and the crystallization starts. This is likely to be higher than the initial coated volume fraction. In the case of seeded growth, it is the volume fraction of the complete amorphous phase, including the seed crystals, at the coating step. The volume fraction $\phi'_{0}$ that has to be coated on top of the half circular seeds can be calculated by
\begin{equation}
    \phi'_{0} = \frac{2Lh_0\phi_0 - \pi r_0^2}{2Lh_0 -\pi r_0^2}.
\end{equation}

Within this modeling picture, we can investigate the effect of geometry (film thickness, crystal spacing and initial size) and global mass transfer between phases (evaporation and growth rate, initial solute amount) on the final morphologies. We can also find process design rules by relating the process-related model parameters to the film formation mechanisms and final morphologies. 

\section{Identification of the possible formation pathways}

Within the framework of the model introduced in the last section, we have identified four different events that may occur during evaporation and crystal growth, which sequence defines the formation pathway. The four events are the termination 
of evaporation (abbreviation E, no solvent left in the liquid film), the termination of crystal growth (abbreviation G, no solute left in the liquid film), crystal impingement (abbreviation I), and the formation of a contact area between crystals and the vapor phase when the crystals touch the film surface (abbreviation T). Among these, crystal impingement is the only event that may not systematically occur, since growth might be insufficient for crystals to reach each other, as will be shown in detail below. These four events are sufficient and necessary to distinguish the possible formation pathways/morphologies that can occur in the model. They may be, in principle, sorted into 4!=24 different sequences. However, certain sequences are non-physical:

\begin{enumerate}
    \item The event of crystals touching the film surface (T) requires either crystal growth or evaporation to be still ongoing processes. Thus, all sequences with T occurring after both events G and E are impossible.
    \item Similarly, crystal impingement cannot take place if crystal growth is already terminated. Therefore, all sequences with event G occurring before event I are not possible. Note that sequences four and nine are possible; however, impingement does not occur: the morphologies resulting from these formation pathways feature pinholes.
    \item When evaporation terminates before crystal growth, a layer of pure solute forms on the substrate. Consequently, there cannot be pinholes in the film and impingement necessarily occurs. This makes sequences with event E occurring before event G, and without event I, not possible.
\end{enumerate}

This gives us eleven possible morphology formation pathways, which are listed in \autoref{tab:Mechanism}. In the following, we investigate in detail for which process and material parameters the various formation pathways take place, and to which morphologies they lead.

\begin{table}[h!]
\caption{Formation pathways:
The 11 formation pathways are listed according to the sequence of possible events E (end of evaporation), G (end of crystal growth), I (crystal impingement / '-' no impingement) and T (crystals touch the film surface). A sketch of the different formation pathways is given in \autoref{fig:finalMor}.}
    \centering
        \begin{tabular}{ |l|l| l| l|l| }\hline
        Number & 1st event & 2nd event & 3rd event & 4th event\\\hline
        1 & E & T & I & G \\
        2 & I & E & T & G \\
        3 & E & I & T & G \\
        4 & G & T & E & - \\
        5 & I & G & T & E \\
        6 & T & E & I & G  \\
        7 & I & T & E & G  \\
        8 & T & I & E & G  \\
        9 & T & G & E & -  \\
        10 & I & T & G & E  \\
        11 & T & I & G & E  \\\hline
        \end{tabular}
\label{tab:Mechanism}
\end{table}

\section{The role of drying rate and crystal number density}\label{sec:BasicDiagram}

\subsection{Identification of the formation pathways depending on $v_e/v_g$ and $L/h_0$}

\begin{figure*}[!ht]
\centering
  \includegraphics[width = 0.95\textwidth]{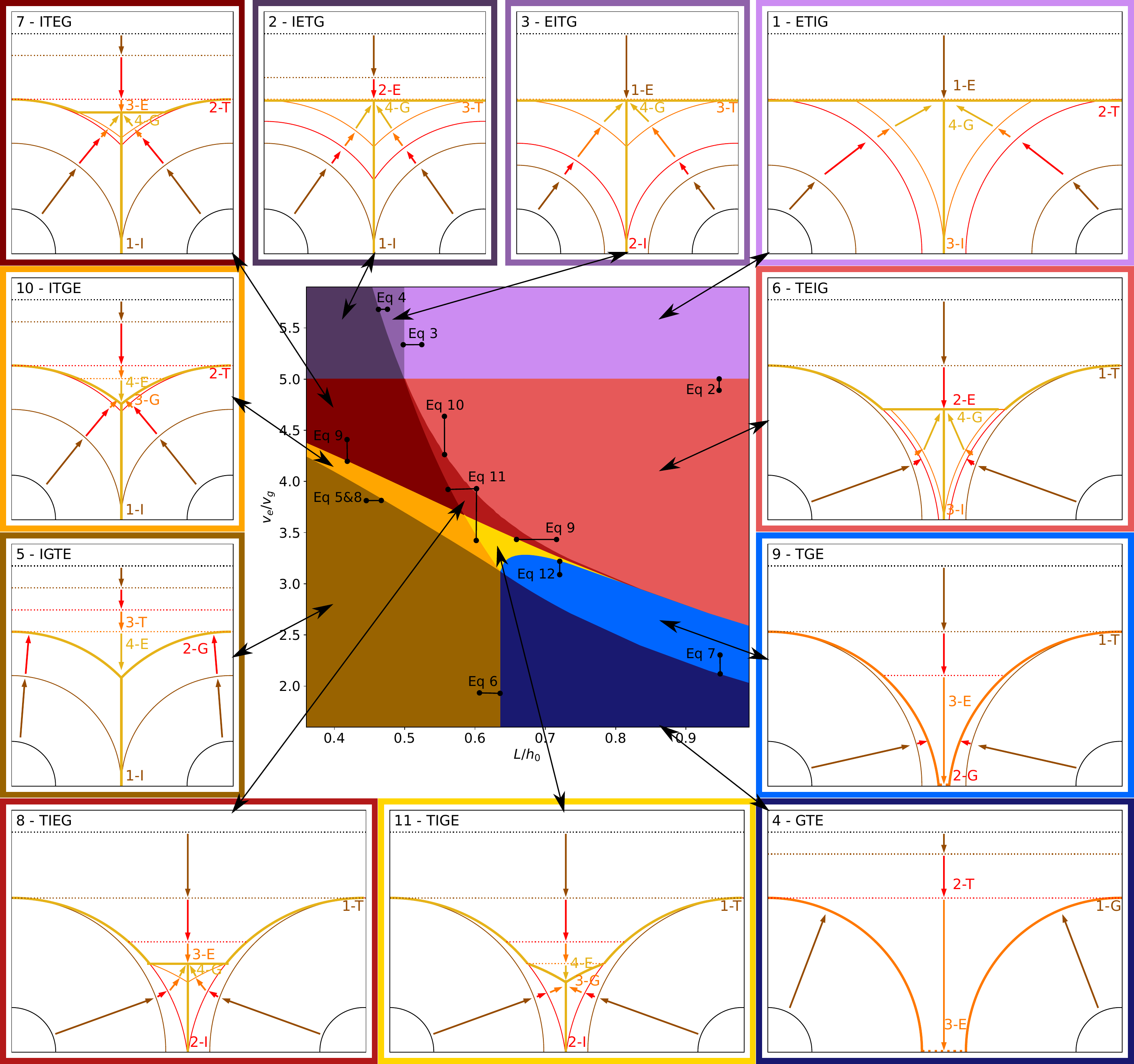}
  \caption{Center: The mapping of the formation mechanisms onto the parameter space. Each formation mechanism is represented by a different color. Each type of final morphology is represented by similar colors (eg. light and dark blue for the morphology including pinholes). The equations from the main text used to distinguish the different sequences are connected to the respective limit. Surrounding: A sketch depicting the formation pathways of the classified morphologies. Only one interface between the crystals is shown. The formation pathways are connected to their place in the mapping. The occurrence of the four formation events, E: end of solvent evaporation, G: end of crystal growth, I: crystal impingement and T: contact between crystalline and vapor phase, are shown. The solid lines depict the extend of the crystalline phase and the dotted line the surface of the  condensed film at the respective time step. The arrows indicate the temporal evolution of the system.}
  \label{fig:finalMor}
\end{figure*}

We aim at determining the boundaries between the different formation pathways depending on the drying rate and the crystal number density. More specifically, it turns out that the ratio of evaporation to growth rate $v_e/v_g$, and of the distance between crystal centers to the initial film height $L/h_0$ are the two related parameters that best describe the existence of the formation pathways in the parameter space. The following section presents the conditions defining the existence domains of the formation pathways. The derivation of all the corresponding equations is provided in the Supporting Information, sections 1-2. \autoref{fig:finalMor} provides an overview of the formation pathways in the 2D-parameter space ($L/h_0$, $v_e/v_g$) at fixed initial volume fraction $\phi_0$ and ratio of initial crystal size to initial film height $r_0/h_0$. 

\subsubsection{Fast evaporation: evaporation finishes before crystals touch the film surface (E $\rightarrow$ T $\rightarrow$ G)}\label{sec:SuperFastEvap}

Formation pathways no. 1, 2, 3: If evaporation is very fast as compared to growth (large $v_e/v_g$), evaporation may come to an end before crystal growth reaches the film surface (E before T). Since T may not occur after both E and G, the sequence of the three events needs to be ETG. For all of these sequences, evaporation ending before any crystal-vapor contact implies that the time for crystal growth up to the film surface is longer than the time for full solvent evaporation. Writing this condition as an equation and rearranging results in 
\begin{equation}\label{equ:fastEvap}
    \frac{v_e}{v_g} > \frac{(1-\phi_0)}{\phi_0 - r_0/h_0}.
\end{equation}
Note that this condition does not depend on $L/h_0$, which implies a horizontal boundary in the 2D parameter space shown in \autoref{fig:finalMor}.

Nevertheless, three different formation pathways are possible, depending on when impingement occurs. Obviously, impingement occurs earlier when the crystals are initially closer to each other. This means that, in general, formation pathways corresponding to 'early' impingement correspond to the lower $L/h_0$ ratios, and those corresponding to 'late' impingement to the higher $L/h_0$ ratios. We begin with the case of 'late' impingement, then consider the case of 'early' impingement and finally discuss the intermediate formation pathway.

ETIG/no.1: in the case of 'late' impingement, termination of evaporation (E) is the first step. As a result, the final state will be completely flat. Following this, the crystals grow up to the surface (T). They impinge (I), and  growth eventually terminates when the film is fully converted to the crystalline state (G). The condition that the crystals first touch the film surface before impinging leads to
\begin{equation}\label{equ:sepFastEvap}
    \frac{L}{h_{0}} > 2\phi_0.
\end{equation}
Note that this condition does not depend on $v_e/v_g$, which implies a vertical boundary in the 2D parameter space in \autoref{fig:finalMor}.

IETG/no.2: in the case of 'early' impingement, crystal impingement is the first event (I). Following this, evaporation terminates (E). Then, the crystals grow up to the film surface (T), and finally, crystal growth terminates when all the film is converted to the crystalline phase (G). This results in a final morphology similar to the previous one, characterized by no roughness and rectangular crystals. The condition that impingement occurs before the termination of evaporation leads to the following equation:

\begin{equation}
    \frac{v_e}{v_g} < \frac{1-\phi_{0}}{\frac{L}{2h_0}-\frac{r_{0}}{h_0}}
\end{equation}

EITG/Nr.3: The final formation pathway corresponding to 'intermediate' impingement is located between the two previous formation pathways in the 2D parameter space (see \autoref{fig:finalMor}). Thereby, evaporation stops (E) before crystals impinge (I). Subsequently, the crystals grow to the surface (T) before crystal growth eventually stops (G).

\subsubsection{Slow evaporation: crystal growth terminates before the crystal touches the surface (G $\rightarrow$ T $\rightarrow$ E)}\label{sec:cryGrowth}

Formation pathways no. 4, 5: If evaporation is very slow as compared to growth (small $v_e/v_g$), crystal growth may terminate while the crystals are still fully covered by solvent (G before T). Since T may not occur after both E and G, the sequence of the three events needs to be GTE. There is a maximal ratio of $v_e/v_g$ for which growth ends before the crystal and the film surface touch each other. It is derived from the comparison between the time it takes to consume all the suitable material by crystal growth, and the time it takes for the liquid-vapor interface to travel down to the height $h_{max}$ (top level of the fully grown crystals). This leads to
\begin{equation}\label{equ:IGTE}
    \frac{v_e}{v_g} < \frac{h_{0}-h_{max}}{h_{max}-r_{0}}.
\end{equation}
The maximum height of the crystals $h_{max}$ depends on crystal geometry and thus on the occurrence of impingement.

GTE-(without I), no. 4: In this formation pathway, growth terminates first (G). Following this, the film surface touches the fully grown crystals due to continued evaporation (T). Finally, the evaporation stops when there is no solvent remaining (E). Impingement of the crystals is not even possible, so that the final morphology consists of two separated half-spherical crystals, with pinholes between them. The fact that the whole solute is contained in not-touching, half-spherical crystals, allows us to derive the first existence condition for this formation pathway:
\begin{equation}\label{equ:pinholes}
    \frac{L}{h_{0}} > \frac{8\phi_0}{\pi}
\end{equation}
Due to the simple spherical geometry, the maximum crystal height can also be derived easily. This provides the second existence condition (on $v_e/v_g$) for this formation pathway as
\begin{equation}\label{equ:GTEI}
    \frac{v_e}{v_g} < \frac{1 - \sqrt{\frac{2L\phi_{0}}{h_0\pi}}}{\sqrt{\frac{2L\phi_{0}}{h_0\pi}} - r_{0}/h_0}.
\end{equation}

IGTE, no. 5: In this formation pathway, corresponding to 'early impingement',  impingement is actually the first event (I). Subsequently, crystal growth ceases when the solute volume is exhausted (G). The evaporation then causes the film surface to come into contact with crystals (T), and finally, the remaining solvent fully evaporates (E). The final morphology is pinhole-free and consists of impinged crystals with spherical caps. Evaluating the crystalline volume in this configuration, and knowing that it is equal to the total solute volume, allows us to calculate $h_{max}$ by (numerically) solving the following equation (compare SI, section 2.3):
\begin{equation}\label{equ:hmax}
    \frac{L}{2}\cdot \sqrt{h_{max}^2 - (L/2)^2} + h_{max}^2\text{arcsin}\left(\frac{L}{2h_{max}}\right) = 2L \cdot  \phi_{0} \cdot h_{0}
\end{equation}
Inserting the value of $h_{max}$ in \autoref{equ:IGTE} provides the upper limit of the ratio of $(v_e)/(v_g)$ for which this formation pathway exists.

\subsubsection{'Intermediate fast' evaporation: evaporation terminates before crystal growth (T $\rightarrow$ E $\rightarrow$ G)}

 For these formation pathways (no. 6, 7, 8) evaporation terminates before crystal growth (E before G). Evaporation is not as fast as compared to the sequences considered in section \ref{sec:SuperFastEvap}, so that the crystals and the film surface have sufficient time to touch before the solvent evaporation ends (TEG sequence). The upper existence limit along the $v_e/v_g$ direction is given by \autoref{equ:fastEvap}, beyond which the formation pathways discussed in section \ref{sec:SuperFastEvap} are active. The lower $v_e/v_g$ limit separating them from the formation pathways at 'intermediate slow' evaporation (see below) has to be solved for numerically. Thereby, we find the values for which the evolution of the amount of solvent
\begin{equation}\label{equ:solvent}
    V_{s}(t) = 2L(1 - \phi_{0})h_0 - v_e\cdot \frac{(h_{0}-r_{0})}{(v_g+v_e)}\cdot 2L -\int_{\frac{(h_{0}-r_{0})}{(v_g+v_e)}}^{t} dt \,(v_e\cdot (2L-4\sqrt{(r_0 + v_gt)^2 - (h_{0}-v_e\cdot t)^2}))
\end{equation}
has a minimum at $V_{s}(t_{min}) = 0$. The origin of this condition is explained in detail in the SI, Sec.~2.4. Here again, we can distinguish three different pathways, depending on when impingement takes place. We discuss the case of 'late' impingement first, then 'early' impingement and finally the intermediate case.

TEIG, no. 6: For this formation pathway with 'late' impingement, the crystals first come into contact with the film surface (T). Next, evaporation terminates (E). Subsequently, the crystals impinge (I), and the crystal growth stops (G). The resulting morphology is similar to the two previous one. To distinguish this region from the remaining one in this set, we use the fact that impingement happens later than the end of evaporation. We, therefore, solve for the ratio of $v_e/v_g$ for which the impingement time is larger than the time $t_s$ for the solvent to be fully evaporated,
\begin{equation}
    t_i = \frac{L/2 - r_{0}}{v_g} > t_s,
\end{equation}
where $t_s$ is computed numerically (see details in the SI, Sec.~2.4).

ITEG, no. 7: In this formation pathway, impingement occurs first (I), followed by the crystals' contact with the (moving) film surface (T). Evaporation eventually ceases (E) earlier than the crystal growth (G). As a result, the shape of the crystals outside of the grain boundaries will no longer be spherical caps, because crystal growth continues only in the part of the crystal surface that remains covered by liquid. Additionally, a flat valley eventually forms in-between the crystals where the film surface stops at the end of evaporation. In order to identify the last limit for the existence of this pathway, we use the fact that impingement happens earlier than the formation of the crystal-vapor interface. This leads to
\begin{equation}\label{equ:I_T}
    \frac{v_e}{v_g} < \frac{2 - L/h_{0}}{L/h_{0}-2r_{0}/h_{0}}.
\end{equation}

TIEG, no. 8: In this sequence with 'intermediate' impingement, crystals first reach the vapor interface (T). This is followed by impingement (I), and finally, evaporation stops (E) a bit earlier than crystal growth (G). The resulting morphology is again similar to the previous ones. 

\subsubsection{'Intermediate slow' evaporation: crystal growth terminates before evaporation (T $\rightarrow$ G $\rightarrow$ E)}

For the last three evaporation pathways (no. 9, 10, 11), evaporation is slower than for the previous sequences. The first event is still the contact between crystals and film surface (T), but this time growth terminates before evaporation (TGE sequence). In the 2D parameter space shown in \autoref{fig:finalMor}, the existence region for the three remaining sequences is limited by \autoref{equ:GTEI}, and \autoref{equ:IGTE} plus \autoref{equ:hmax} (from below) as well as \autoref{equ:solvent} (from above). Again, the three cases of 'late', 'early' and 'intermediate' impingements have to be considered.

TGE-(without I), no. 9: For this sequence, the crystals first come into contact with the vapor interface (T). Subsequently, crystal growth stops (G), and finally, evaporation terminates (E). There is no impingement. The final morphology consists of distorted half-spheres with pinholes in between. The condition for the absence of impingement is also used to identify the existence region of this formation pathway, by solving numerically
\begin{equation}
    \pi (r_{0} + (h_{0}-r_{0})/(v_g+v_e)\cdot v_g)^2 +
    \int_{(h_{0}-r_{0})/(v_g+v_e)}^{t} dt\, (4\cdot r(t)\text{arcsin}\left(\frac{h_{0} - v_e\cdot t}{r(t)}\right) \cdot v_g) = 2L(1 - \phi_{0})h_0
\end{equation}

ITGE, no. 10: In this case, impingement happens first (I). Afterward, the crystals touch the film surface (T) and continue growing. The resulting crystal shape outside of the grain boundaries deviates from the spherical cap, as before. Now, crystal growth (G) stops earlier than evaporation (E).  As a result, in this case, there is a trench instead of a flat valley in between crystals. Similar to the formation pathway ITEG, the difference we exploit to find the last existence limit for this sequence is that impingement happens earlier than the creation of a crystal-vapor interface. Here again, this leads to \autoref{equ:I_T}.

TIGE, no. 11: For the final formation pathway, the crystals grow to the vapor interface (T) and then impinge (I). Crystal growth terminates (G) earlier than evaporation (E). Since crystal growth continues after the contact of the crystals with the vapor phase, the crystal shape developing before impingement deviates from being perfectly spherical. Due to impingement, there are no pinholes in this film. Between the crystals, there is a trench.

\subsection{Morphologies and process design rules}

In summary, there are eleven possible formation pathways within the model's hypotheses, leading to four types of morphology. The first morphology arises when the ratio of drying rate to growth rate is high. For many slowly crystallizing systems, that may already be the case for medium or even low drying rates. The film is pinhole-free, perfectly flat, and consists of rectangular crystals. This morphology is created by the formation mechanisms 1, 2, and 3 (ETIG, IETG, EITG), corresponding to the purple regions in \autoref{fig:finalMor}. For the formation pathways 6, 7, 8 (TEIG, ITEG, TIEG) at 'intermediate high' $v_e/v_g$, the film is pinhole-free but rough, with flat regions between the crystal domes. These morphologies are marked in red in \autoref{fig:finalMor}. For the formation pathways 5, 10, 11 (ITGE, IGTE, TIGE) at low $v_e/v_g$ and low $L/h_0$ (high seed density), the film is still pinhole-free and rough, but with trenches between the crystal domes. These morphologies are marked in yellow in \autoref{fig:finalMor}. Finally, for the formation pathways 4 and 9 (GTE, TGE) at low $v_e/v_g$ and high $L/h_0$ (low seed density), pinholes are present in the films between fully separated crystals. These morphologies are marked in blue in \autoref{fig:finalMor}.

\begin{figure}[h!]
\centering
  \includegraphics[width = 0.45\textwidth]{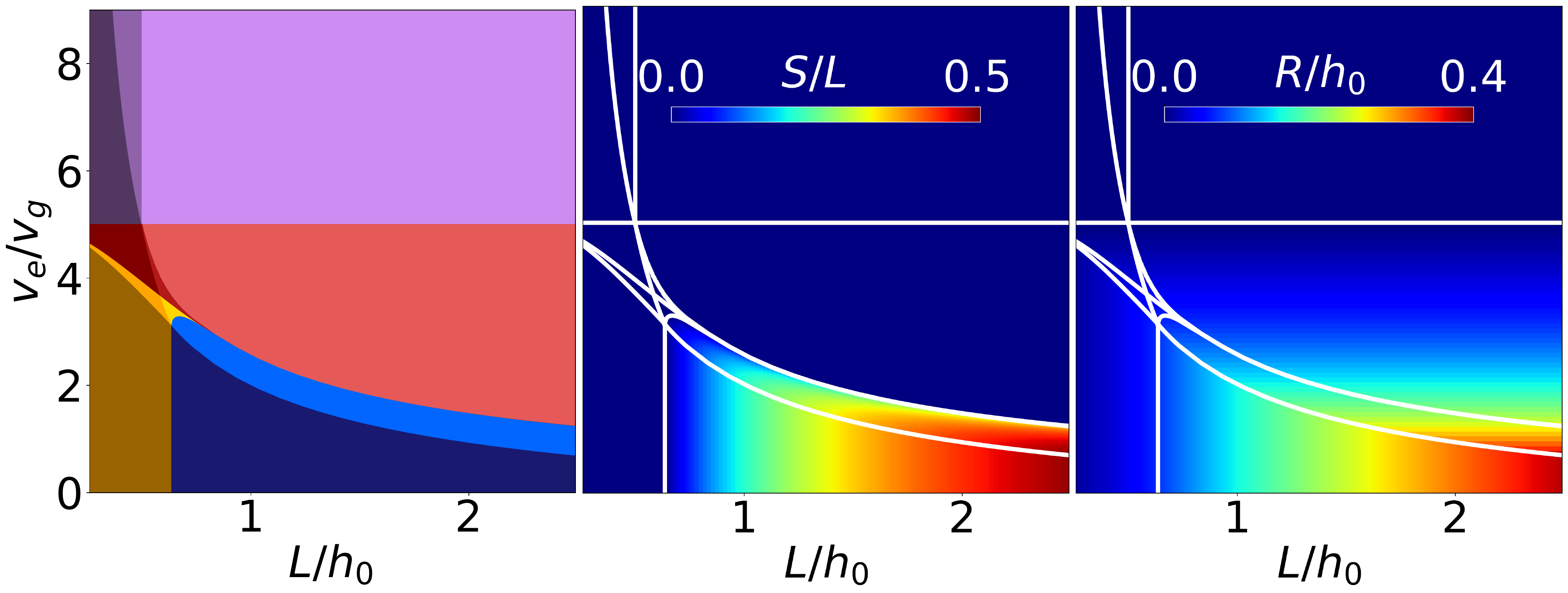}
  \caption{Analysis of substrate coverage and film roughness. Left: mapping of the formation pathways and morphology types based on the evaporation-to-growth rate ratio and the system size. Each formation mechanism is represented by a different color. Each type of final morphology is represented by similar colors. Center: The amount of uncovered substrate (S) between the crystals. Right: Roughness of the final morphology R, calculated as the ratio between the highest point of the film and the average film thickness. The boundaries between the formation pathways are indicated with white lines. For all the figures, the initial volume fraction of crystalline material is $\phi_0 = 0.25$, and the initial ratio of crystal size to film height is $r_0/h_0 = 0.1$ (same parameters as in \autoref{fig:finalMor}).
  }
  \label{fig:RoughUnc}
\end{figure}

We now turn to the evaluation of substrate coverage and film roughness in the dry film, depending on the film formation pathways (\autoref{fig:RoughUnc}). The existence regions of the different formation mechanisms with respect to $v_e/v_g$ and $L/h_0$ are shown on the left. In the center, the amount of uncovered substrate is depicted. The roughness of the film, as defined by the ratio of the highest point in the dry film to the average film height, is shown on the right.

There are pinholes in the film only for the formation pathways 4 and 9. Thereby, the amount of uncovered substrate increases with increasing $L/h_0$. The change in the amount of uncovered substrate between sequences 9 and 6 (light red) is quite drastic. This corresponds to the transition between pinholes, if crystal growth ceases earlier than evaporation, and an (infinitely) thin crystalline layer between the crystals, if evaporation ceases earlier than crystal growth.

The roughness of the dry films increases with increasing $L/h_0$ and decreasing $v_e/v_g$. For higher evaporation rates (lower growth rates), the crystals cannot grow far in the vertical direction, hence the roughness is lower. For a larger value of $L/h_0$ there is more crystalline material distributed on each crystal, allowing for a higher roughness. The film is expected to be perfectly flat only for sufficiently high $v_e/v_g$ ratios.

From a practical perspective, the substrate should be fully covered with perovskite. Consequently, the region of the configuration space that should be strictly avoided is the blue region 
(sequences 4 and 9) appearing at low $v_e/v_g$ and medium to high $L/h_0$. That is one reason why a high seed density (corresponding to low $L/h_0$) or a high evaporation rate (high $v_e/v_g$) is advantageous for depositing the active layer of a high-performing solar cell~\cite{ding_material_2017,li_vacuum_2016,huang_gas-assisted_2014}. Note that perovskite materials often crystallize so fast that the desirable $v_e$ may sometimes not be accessible experimentally. In such cases, using seeded growth or favoring heterogeneous nucleation on the substrate with a high seed number density (down-left part of the parameter space) might be an alternative.

On top of the mechanisms discussed in this work, further physical processes may take place in real systems. For instance, fast evaporation can lead to supersaturation in the liquid phase and thus trigger homogeneous nucleation in the film. This strongly influences the film morphology~\cite{majewski_simulation_2025}. Since a high evaporation rate promotes supersaturation and thus nucleation, there is a large number of crystals. The crystals arrange in such a way that the final film is relatively flat, which is positive. However, for device performance, a small amount of large crystals is preferred, because crystal boundaries lead to charge carrier recombination.
To obtain a nearly smooth film surface with a small number of crystals (and therefore interfaces) and a low evaporation rate, an alternative formation pathway is to use a high seeding density (sequence 5, dark yellow). The roughness of the film also decreases with increasing seeding density (compare \autoref{fig:RoughUnc}).

\section{The impact of the model parameters}\label{sec:phi_r_h}

\begin{figure}[h]
\centering
  \includegraphics[width = 0.45\textwidth]{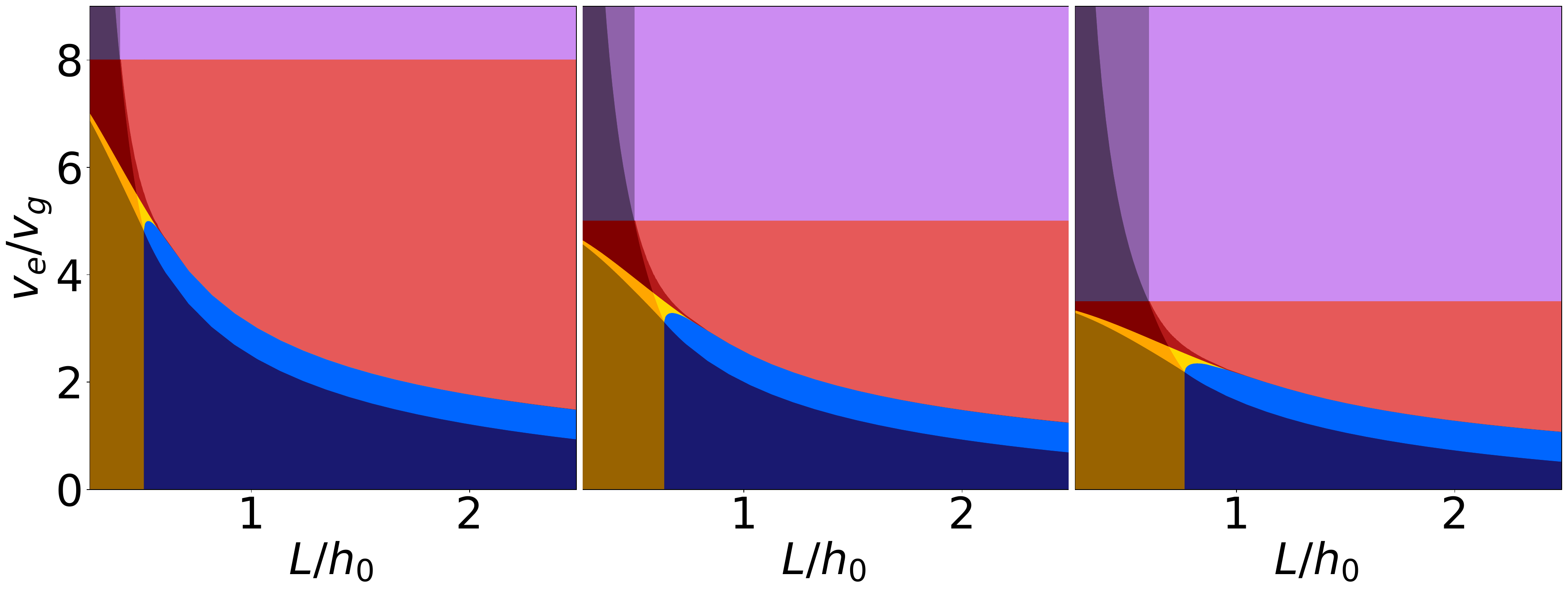}
  \caption{Mapping of the formation pathways and morphology types depending on initial volume fraction ($\phi_{0}$). $\phi_{0}$ increases from left to right ($\phi_{0}$ = 0.2, 0.25, 0.3). The initial ratio of crystal size to film height is $r_0/h_0 = 0.1$. Each formation mechanism is represented by a different color. Each type of final morphology is represented by similar colors.
  }
  \label{fig:Phi}
\end{figure}

In this section, we investigate the effect of changing the two remaining relevant process parameters, the initial solute volume fraction $\phi_0$ and the ratio of the initial crystal size to the initial film height $r_0/h_0$. We first consider the effect of varying the initial volume fraction. To this end, three plots depending on $v_e/v_g$ and $L/h_0$ with different initial volume fractions are shown in \autoref{fig:Phi}. The roughness and the amount of uncovered substrate are shown in the SI, section 3. The central plot corresponds to the parameters already used in \autoref{sec:BasicDiagram}. Upon changes of the initial volume fraction, the pattern of the existence domains for the 11 formation pathways remains qualitatively unchanged. However, the exact location of the boundaries between the sequences shifts in response to changes in the initial volume fraction. A comparison of the three boundary representations reveals two notable effects: First, the boundaries shift towards lower ratios of $v_e$/$v_g$ as the initial volume fraction increases. Second, the vertical boundaries shift towards larger $L/h_0$.

The first effect can be understood qualitatively by considering that a higher initial solute volume fraction for the same initial film height means that the amount of solvent is lower. Consequently, evaporation does not need to be as fast to have the same effect as it would for a lower initial solute volume fraction. This leads to a downward shift of the boundaries for higher $\phi_{0}$. The second effect can be understood as follows: the interface position between a film with and without pinholes (compare \autoref{equ:pinholes}) and the interface at high evaporation rates (compare \autoref{equ:sepFastEvap}) are directly proportional to $\phi_0$. For the interface at medium ratios of evaporation and growth rates, the effect is indirect. \autoref{equ:I_T} is independent of $\phi_0$, but the upper and lower limit for the evaporation to growth rates shifts to lower values (as explained before), and therefore these sequences seemingly shift to higher ratios of $L/h_0$. Intuitively, one can think of the following:
Given a higher initial volume fraction, there is more solute material in the system, resulting in larger crystals in the final state. This implies that larger distances between the crystals are required to produce a final film with pinholes/without impingement. This explains the shift towards larger $L/h_0$ for low $v_e/v_g$. For high ratios of $v_e/v_g$, the main difference is the increase in the (average) final film thickness, which increases the time until the crystal-air interface forms. 

From a practical point of view, using a high evaporation rate may be experimentally complicated. Increasing the initial volume fraction allows for a reduction in the evaporation rate required to get a smooth film surface. Increasing the initial volume fraction also requires a lower seed density, at a low evaporation rate, to obtain a fully covered film.

\begin{figure}[h]
\centering
  \includegraphics[width = 0.45\textwidth]{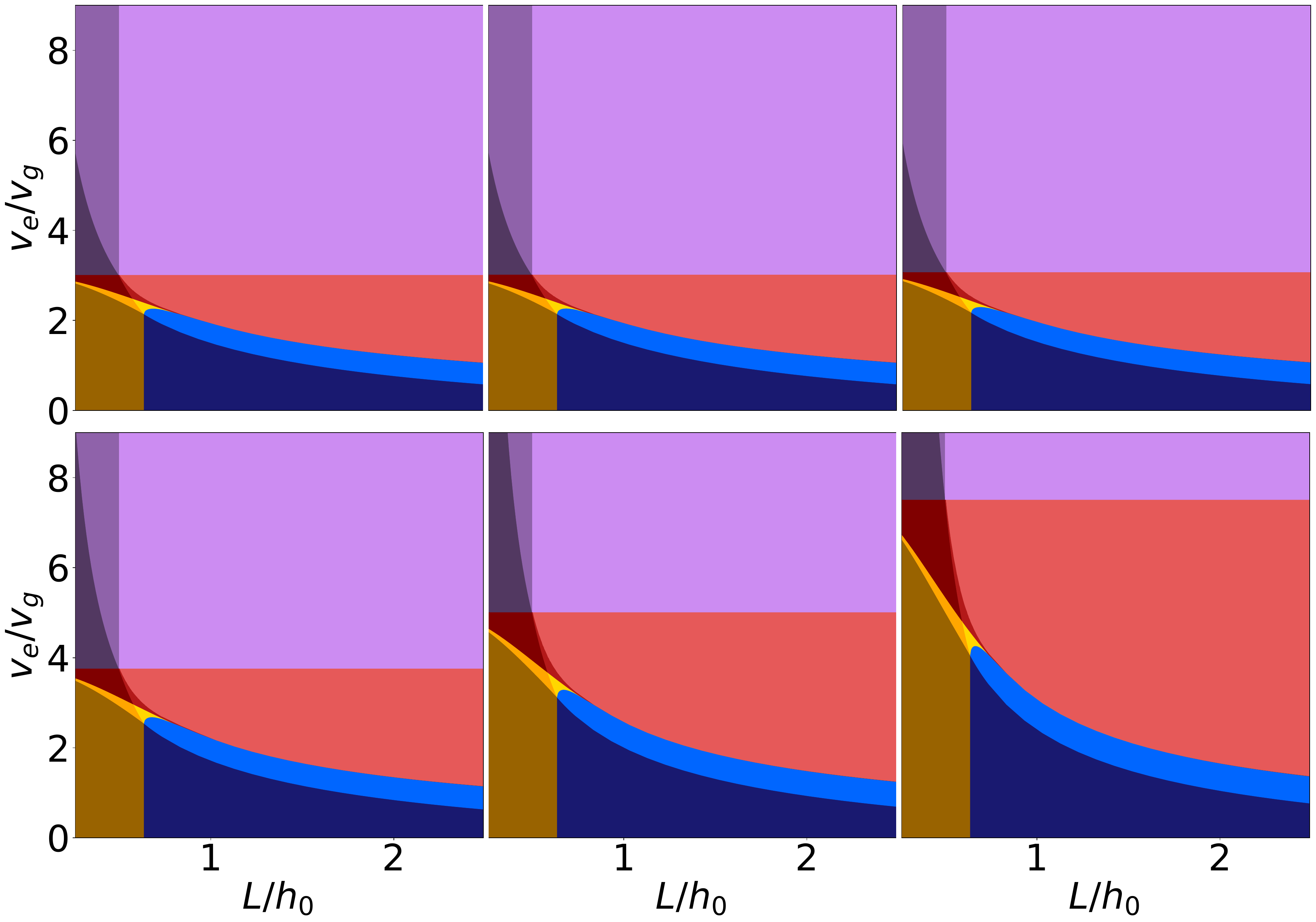}
  \caption{How the mapping of the formation pathways and morphology types changes with initial crystal size/initial film height variation. The ratio $r_0/h_0$ increases from left to right and top to bottom (first row: $r_0/h_0$ = $5\cdot10^{-5}$, $5\cdot10^{-4}$, $5\cdot10^{-3}$ second row: $r_0/h_0$ = 0.05, 0.1, 0.15). The initial volume fraction is $\phi_0 = 0.25$. For very small ratios of $r_0/h_0$ the effect on the representation is negligible.}
  \label{fig:Rini}
\end{figure}

Finally, we investigate the impact of the ratio of initial crystal size to initial film height $r_0/h_0$. Domain existence representations corresponding to six different ratios $r_0/h_0$ are presented in \autoref{fig:Rini}. The roughness and the amount of uncovered substrate are shown in the SI, section 3. Here again, the central plot in the second line corresponds to the parameters already used in \autoref{sec:BasicDiagram}, and the pattern of the existence domains remains qualitatively unchanged. A comparison of the boundary representations reveals three notable features: First, for low ratios of $r_{0}/h_{0}$, the boundaries between the existence domains remain (nearly) unchanged. Second, the boundaries between existence domains remain unchanged in the horizontal direction for any ratio. Third, increasing the ratio $r_{0}/h_{0}$ leads to a shift of the boundaries towards higher ratios of $v_e$/$v_g$.

This can be understood as follows: on the one hand, decreasing the initial crystal sizes increases the time the crystal needs to grow to a given size for the same crystal growth rate. To reach this size in the same amount of time, the growth rate must be larger, which means smaller $v_e/v_g$ ratios. This is the reason why the boundaries shift to smaller 
$v_e$/$v_g$ for smaller initial crystal sizes. This effect saturates when the initial crystal size becomes very small compared to the initial film thickness. On the other hand, the final crystal volume is the relevant parameter in determining the boundaries in the x-direction at low and high ratios of $v_e$/$v_g$. Since the initial solute volume fraction is calculated on the whole amorphous phase (including crystals), the change of crystal size does not change the amount of crystalline material (see \autoref{sec:model}). Hence, the boundaries do not move in the x-direction. 

From a practical perspective, in the case of dominant heterogeneous nucleation at the substrate, the initial crystal size can be viewed as a simplified model for the critical radius, as described in the classical nucleation theory~\cite{dunlap-shohl_synthetic_2019,er-raji_insights_2023}. According to this theory, there is a stable crystal size determined by a competition between interfacial and bulk energies. For perovskite solar cells, the system is in the regime of very low ratios of $r_{0}/h_{0}$. Hence, changing the crystal size will not change the formation pathway. A slight increase in the initial film height corresponds to a movement to the left in the representation. Notably, the necessary ratio of evaporation to growth rate does not change. 

For larger initial crystal sizes, we see that a larger critical radius requires a higher evaporation rate (or decreased growth rate) to produce a smooth film. Hence, it appears that a smaller initial crystal radius is favorable for the fabrication of smooth films because it allows for a reduced evaporation rate. An intuitive understanding of a change of the initial film height in this regime is complicated in this framework and is discussed in the SI, section 4. In the case of controlled patterning with seeds, the seed size can be controlled. In this context, decreasing the seed size can be seen as a good way to favor the fabrication of smooth films, because the processing window in terms of evaporation rate becomes larger.

\section{Summary and Outlook}

A model was developed to describe the growth of perovskite crystals from the substrate in a drying solution. This allows for the investigation of how the interplay between evaporation and crystal growth determines the dry film morphology. We identified four steps in the film formation: the impingement of the crystals, the formation of a contact line between the crystalline and vapor phase, the end of crystal growth, and the end of evaporation. The sequence of these events defines eleven possible formation pathways, leading to four different types of dry film morphologies that can be distinguished from each other, particularly regarding substrate coverage and film roughness.  
We found two possible ways to adjust processing parameters to obtain a final film with a smooth surface and full substrate coverage. We confirmed that the well-known approach using fast evaporation~\cite{majewski_simulation_2025} is a very effective way. More precisely, the evaporation rate has to be high enough compared to the crystal growth rate. This is the fundamental reason why the fast crystallizing metal-halide perovskites often require very fast drying. Another approach is to ensure a sufficiently high seed density at the substrate, which allows the use of low evaporation rates.

The proposed model has the advantage of being simple enough to be analytically tractable, and allows for the consideration of very basic physical properties of crystallizing film formation. While reality is more complex, we expect our findings to remain qualitatively valid even with further refinements and additional physics taken into account, such as surface tension effects, description of possible diffusion limitation, non-constant growth and evaporation rates, and generalization to 3D (see SI, chapter 8). However, other processes, such as further nucleation during drying or multi-step crystallization through solid-state intermediates, could strongly affect the results of the model. At some point, the problem becomes so complex that simulations become necessary. In order to investigate the impact of these effects, the PF framework proposed by our group \cite{ronsin_phase-field_2022}, which already describes evaporation, nucleation, growth and diffusion properties in detail, will be extended in the future to include dewetting effects and solid state precursors.

\medskip

\textbf{Data Availability}

The implementation of the equations of the model and the simulation data are publicly available at DOI: 10.5281/zenodo.16673320.

\medskip
\textbf{Acknowledgements} \par
We thank Kai Segadlo for helpful discussions and acknowledge financial support from the Deutsche Forschungsgemeinschaft (DFG) via the Perovskite SPP2196 programme (Project No. 506698391), the European Commission (H2020 Program, Project 101008701/EMERGE), and the Helmholtz Association (SolarTAP Innovation Platform).

\medskip


\newpage

\vspace{1em}
\sffamily
\begin{tabular}{p{17.5cm} }

\noindent\LARGE{\textbf{Morphology Formation Pathways in Solution-Processed Perovskite
Thin Films - Supporting Information}} \\
\vspace{0.3cm} \\

\noindent\large{Martin Majewski,$^{\ast}$\textit{$^{a}$} Olivier J.J. Ronsin,\textit{$^{a}\ddag$} and Jens Harting\textit{$^{a,b}$}} \\

\vspace{0.3cm} \\

\noindent\large{\textit{$^{a}$Helmholtz Institute Erlangen-Nürnberg for Renewable Energy (HIERN), Forschungszentrum Jülich GmbH}} \\
\noindent\large{\textit{$^{b}$Department of Chemical and Biological Engineering and Department of Physics, Friedrich-Alexander-Universität Erlangen-Nürnberg, Fürther Straße 248, 90429 Nürnberg, Germany}} \\

\end{tabular}

 \vspace{0.6cm}

\renewcommand*\rmdefault{bch}\normalfont\upshape
\rmfamily
\section*{}
\vspace{-1cm}

\section{Introduction}

In the following, an analytical model is created that can explain the different morphologies that are possible with a patterned substrate with two placed crystals and no further nucleation.

The model is based on the following assumptions:
\begin{itemize}
    \item Direct crystallization is considered only, no SSP or intermediates are present in the system
    \item Only two dimensions are considered
    \item We start initially with two spherical crystals, equidistantly placed
    \item No further nucleation is considered
    \item The growth rate of the crystals (1D interfacial velocity) is constant and drops to zero when no solute in the liquid film is left or the liquid film is consumed completely.
    \item The crystals grow isotropically with a growth rate $v_g$ unless there is no solute anymore in the liquid film, the crystal surface is in direct contact with the air or another crystal.
    \item The evaporation rate ( $v_e = \frac{\text{d(height of the condensed film)}}{\text{dt}}$) is constant and drops to zero when there is no solvent left
    \item Solvent evaporates at the liquid-vapor interface, not at the solid-vapor interface.
\end{itemize}

\begin{figure}[h!]
    \centering
    \includegraphics[width = 0.8\textwidth]{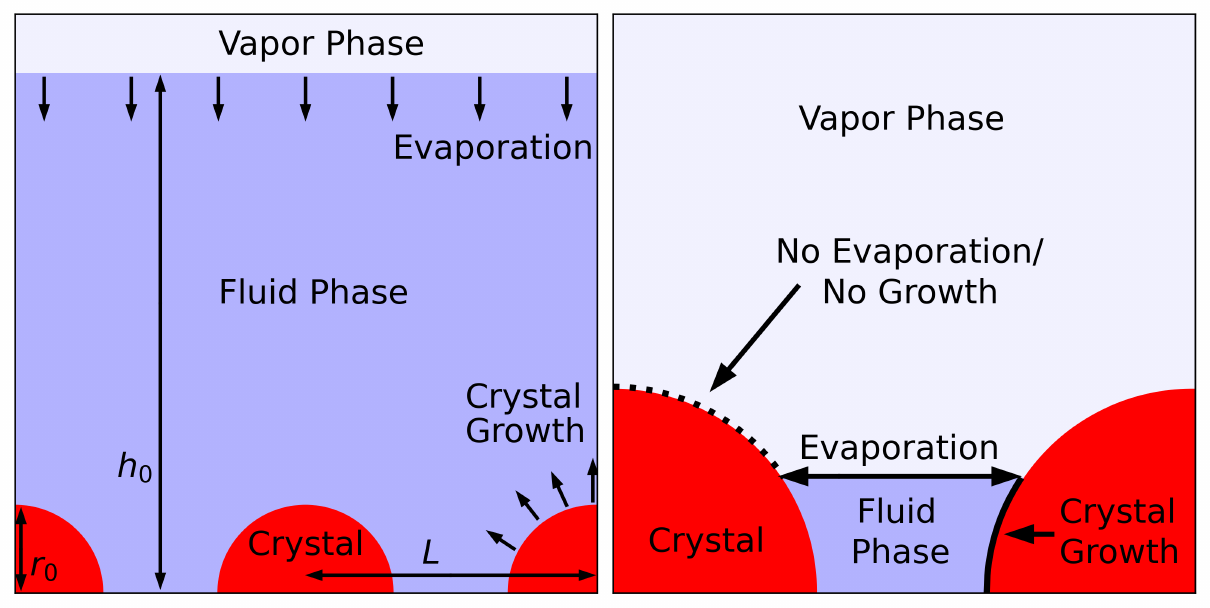}
    \caption{Initial setup. The crystalline (red), fluid (dark blue) and vapor (light blue) phases are shown. L is the distance between the centers of the crystals, which is half of the width of the simulation box. The initial crystal radius $r_{ini}$ and the initial wet film height $h_{ini}$ are visualized.}
    \label{fig:intialSetup}
\end{figure}

\subsection{Initial State}

$L$ is the distance between the crystal centers, hence the box width is $2L$. The initial film height is $h_{0}$, the radius of the two initially placed crystals is $r_{0}$, and the initial solute volume fraction is $\phi_{0}$. The growth rate of the crystals is called $v_g$, and the evaporation rate of the solvent is called $v_e$.

The solute volume $V_{c,tot}$ is constant in time and can be calculated in the initial state via:
\begin{equation}\label{equ:Vtot}
    V_{c,tot} = 2L \cdot  \phi_{0} \cdot h_{0}
\end{equation}
The initial amount of solvent $V_{s,tot}$ is therefore
\begin{equation}
    V_{s,tot} = 2L(1 - \phi_{0})h_0 
\end{equation}
\subsection{Dynamics}

The average height $h_{f}$ of the dry film is 
\begin{equation}\label{equ:Hfinal}
    h_{f} = V_{c,tot}/(2L) = h_{0}\cdot\phi_{0} 
\end{equation}

The time $t_{max}$ it takes to evaporate the solvent completely is between maximally (if evaporation continues till the substrate)
\begin{equation}
    t_{max} = h_{0}/v_e
\end{equation}
for a setup with pinholes and $t_{min}$ minimal
\begin{equation}\label{equ:tmin}
    t_{min} = (h_{0}-h_{f})/v_e
\end{equation}
for a final film that is completely flat.

\section{Possible Final Film Structures}

In the following, the classification described in sec 2 of the main text, is derived analytically. The considered configuration space is the ratio of growth to evaporation rate ${v_e}/{v_g}$ and height of the initial wet film compared to the distance between the crystals $L/h_{ini}$.

\subsection{Evaporation is finished when the crystal touches the surface (E$\rightarrow$ T, Sequence Nr. 1,2,3), purple}

Evaporation is finished before the crystal reach the liquid film surface:
\begin{equation}
    t_{min} < (h_{f} - r_{0})/v_g
\end{equation}
where $r_0$ is the initial crystal radius. Using \autoref{equ:tmin} this leads to
\begin{equation}
    \frac{v_e}{v_g} > \frac{h_{0} - h_{f}}{h_{f}-r_{0}} = \frac{h_0(1-\phi_0)}{h_0\phi_0 - r_0}
\end{equation}

There we can distinguish three cases.

ETIG/Nr.1. The crystals grow up to the surface first and then impinge 
\begin{equation}
    h_{f} - r_{0} < \frac{L}{2} - r_{0}
\end{equation}
and therefore (using \autoref{equ:Hfinal}) this phase is restricted to
\begin{equation}
    \frac{L}{h_{0}} > 2\phi_0
\end{equation}

IETG/Nr.2: The evaporation is finished earlier than the event of impingement:

Time for impingement:
\begin{equation}\label{equ:ti}
    t_i = \frac{L/2 - r_{0}}{v_g}
\end{equation}
The time for evaporation is $t_{min}$. We solve for $t_{min} < t_i$:
\begin{equation}
    \frac{v_e}{v_g} < \frac{h_{0}- h_{f}}{L/2-r_{0}}
\end{equation}

EITG/Nr.3 is placed in-between these two domains.

\subsection{Crystal growth terminates before the crystal touches the surface (G $\rightarrow$ T, Sequence Nr. 4,5), dark blue and dark yellow} \label{sec:growth}

There we can distinguish two cases depending on whether the crystals impinge in the final state (IGTE, Nr.5) or not (GTEI, Nr. 3). 

These two phases are separated by: 
\begin{equation}\label{equ:SepRedBlue}
    L/2 = r_{f}
\end{equation}
where $r_{f}$ is the maximal extension of the crystal:
\begin{equation}\label{equ:Rfin}
    r_{f} = r_{0} + v_g\cdot t_{growth}
\end{equation}
with $t_{growth}$ being the time for which all solute is consumed by growth. This time can be calculated from
\begin{equation}\label{equ:maxGrowth}
    V_c(t_{growth}) = V_{c,tot}
\end{equation}
where $V_{c,tot}$ is the total amount of solute from \autoref{equ:Vtot} and $V_c(t)$ is the time dependent volume of the two (half) crystals:
\begin{equation}\label{equ:Vcry}
    V_c(t) = r(t)^2\cdot\pi = (r_{0} + v_gt)^2\cdot\pi
\end{equation}

Inserting \autoref{equ:Vcry} into \autoref{equ:maxGrowth} and using \autoref{equ:Vtot}, leads to
\begin{equation}\label{equ:tgrowth}
    t_{growth} = \frac{1}{v_g}\left( \sqrt{\frac{2h_{0}\,L\phi_{0}}{\pi}}-r_{0} \right) 
\end{equation}

Inserting this into \autoref{equ:Rfin} and \autoref{equ:SepRedBlue} (and neglecting the option of $L = 0$) results in:
\begin{equation}
    \frac{L}{h_{0}} = \frac{8\phi_0}{\pi}
\end{equation}

GTE, Nr.4: When the crystals do not impinge (dark blue), the condition for growth ending before the crystal touches the surface reads:
\begin{equation}
    r(t_{growth}) < h(t_{grwoth})
\end{equation}
which can be rewritten as:
\begin{equation}
    r_{0}+v_g\cdot t_{growth} = h_{0}-v_e\cdot t_{growth}
\end{equation}
where $t_{growth}$ is calculated by \autoref{equ:tgrowth}. Inserting and rearranging results in
\begin{equation}
    \frac{v_e}{v_g} < \frac{h_{0} - \sqrt{\frac{h_{0}\,2L\phi_{0}}{\pi}}}{\sqrt{\frac{h_{0}\,2L\phi_{0}}{\pi}} - r_{0}}
\end{equation}

IGTE, Nr. 5: In the situation where the crystal impinge, we have (dark yellow):
\begin{equation}
    r(t_{growth}) \geq L/2
\end{equation}

To get an upper limit for $v_e/v_g$ we need to calculate the Area of the grown crystals. The height of the grown crystals $h_{max}$ is connected to the time of crystal growth $t_{growth}$ by
\begin{equation}
    h_{max} = r_{0} + v_g \cdot t_{growth}
\end{equation}

\begin{figure}[h!]
    \centering
    \includegraphics[width = 0.7\textwidth]{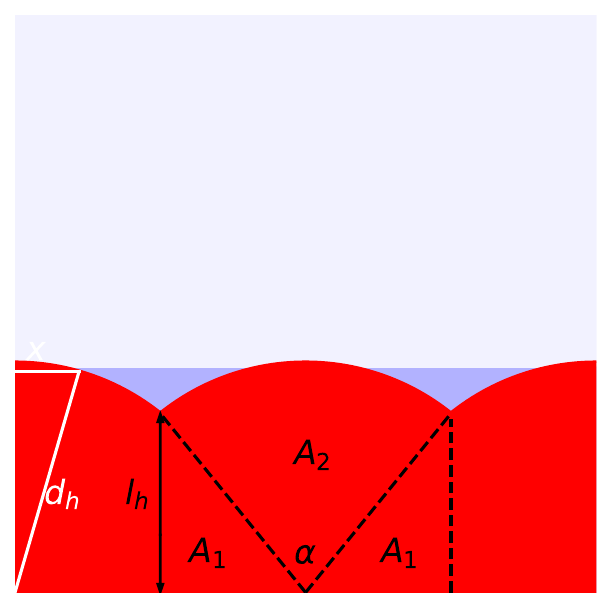}
    \caption{Intermediate stage. The crystalline (red), fluid (light blue) and vapor (dark blue) phases are shown. Shown are the areas $A_1$ and $A_2$, the lengths $l_h$, $h_{max}$ and $x$, and the angle $\alpha$}
    \label{fig:RedPhaseStructure}
\end{figure}

Let the height of the crystals at the grain boundaries be $l_h$. The area of one crystal consists of a circular segment $A_2$ and two triangles $A_1$ (see \autoref{fig:RedPhaseStructure} for definition of $A_1$ and $A_2$). $l_h$ can be calculated in general as:
\begin{equation}
    l_h = \sqrt{h_{max}^2 - (L/2)^2}
\end{equation}
so that the size of the two triangles (together) is
\begin{equation}\label{equ:A1}
    A_1 = \frac{L}{2}\cdot \sqrt{h_{max}^2 - (L/2)^2}
\end{equation}
The angle of the circular segment is 
\begin{equation}
    \alpha = 2\text{arcsin}\left(\frac{L}{2h_{max}}\right)
\end{equation}
so that the size of the area is then
\begin{equation}\label{equ:A2}
    A_2 = h_{max}^2\text{arcsin}\left(\frac{L}{2h_{max}}\right)
\end{equation}

Gathering \autoref{equ:A1} and \autoref{equ:A2} the crystalline volume in the final state $V_{cryst}$ is
\begin{equation}
    V_{cryst} = 2(A_1 + A_2) = \frac{L}{2}\cdot \sqrt{h_{max}^2 - (L/2)^2} + h_{max}^2\text{arcsin}\left(\frac{L}{2h_{max}}\right)
\end{equation}

At the end of the crsytallization process, we have $V_{cryst} = V_{c,tot}$. Using \autoref{equ:Vtot}, $h_{max}$ can be found by solving numerically 
\begin{equation}
    2(A_1 + A_2) = \frac{L}{2}\cdot \sqrt{h_{max}^2 - (L/2)^2} + h_{max}^2\text{arcsin}\left(\frac{L}{2h_{max}}\right) = 2L \cdot  \phi_{0} \cdot h_{0}
\end{equation}

The condition for growth finishing before the crystal touches the surface corresponds to the top of the crystal reaching the position $h_{max}$ before the film surface:
\begin{equation}
    \frac{h_{max}-r_0}{v_g} = \frac{h_0 - h_{max}}{v_e}
\end{equation}
which we rewrite as:
\begin{equation}
    \frac{v_e}{v_g} < \frac{h_{0}-h_{max}}{h_{max}-r_{0}}
\end{equation}

\subsection{The crystals touch the surface.}\label{sec:T}

For all the other cases neither crystal growth nor evaporation is finished when the film surface touches the crystals. The time $t_{T}$ of for the first contact between crystal and film surface is defined by calculating the crystal radius $r_T$ at $t_{T}$ via two different ways :
\begin{equation}
    r_T = r_{0} + v_gt_{T} = h_{0} - v_et_{T}
\end{equation}
and therefore
\begin{equation}\label{equ:Tmeet}
    t_{T} = (h_{0}-r_{0})/(v_g+v_e)
\end{equation}

The volume of the crystals at the time when they com in contact with the film surface is:
\begin{equation}\label{equ:VT}
    V_{T} = \pi (r_{0} + t_{T}\cdot v_g)^2
\end{equation}

Once the crystal and film surfaces have touched each other, growth can terminate either before or after evaporation. To investigate this transition, we can have a look at the solvent in the system: 

The amount of solvent left at $t_{T}$ is 
\begin{equation}
    V_{s,T} = V_{s,tot} - v_e\cdot t_{T}\cdot 2L
\end{equation}

\begin{figure}[h!]
    \centering
    \includegraphics[width = 0.7\textwidth]{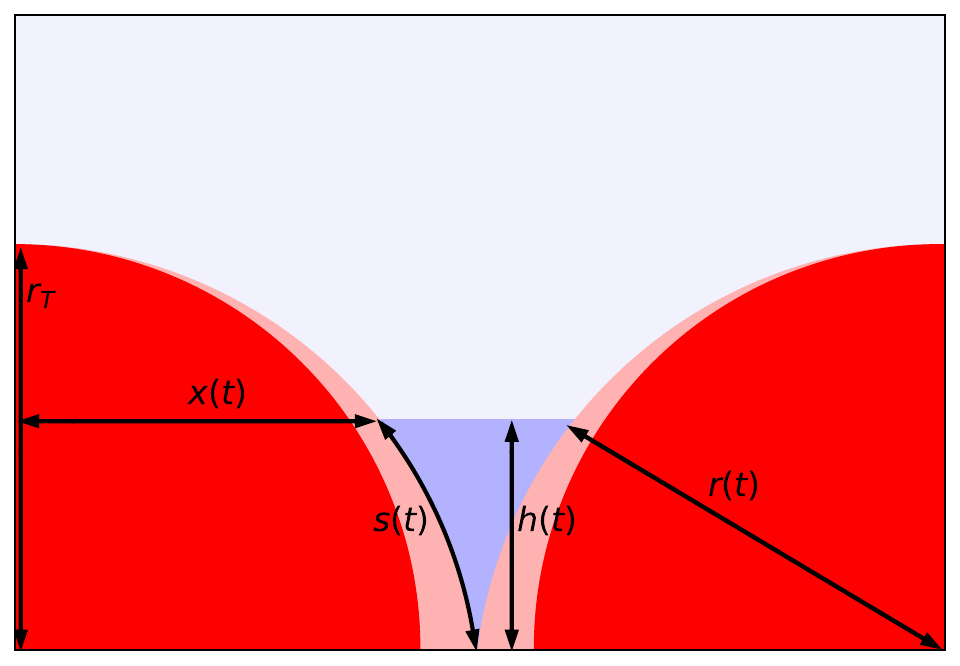}
    \caption{Sketch of an intermediate stage. The crystalline (light red), fluid (light blue) and vapor (dark blue) phases are shown. A circle is displayed in dark red inside the crystal for comparison. Shown are $x(t)$, $r_T$, $r(t)$, $h(t)$ and $s(t)$. }
    \label{fig:RedPhaseStructure}
\end{figure}

The amount of solvent reduces after $t_{T}$ according to
\begin{equation}\label{equ:solvent}
    V_{s}(t) = V_{s,T} - \int_{t_{T}}^{t} dt \,(v_e\cdot (2L-4x(t)) = 2L(1 - \phi_{0})h_0 - v_e\cdot t_{T}\cdot 2L - \int_{t_{T}}^{t} dt \,(v_e\cdot (2L-4x(t)))
\end{equation}
where $x(t)$ is half of the width of a crystal at the height of the liquid-vapor interface (compare \autoref{fig:RedPhaseStructure})
\begin{equation}\label{equ:sqrt}
    x = \sqrt{r^2(t) - h^2(t)} = \sqrt{(r_0 + v_gt)^2 - (h_{0}-v_e\cdot t)^2}
\end{equation}

Note that the calculation of $V_s$ bases on the size of the liquid-vapor interface only, \autoref{equ:solvent} holds, no matter what happens regarding impingement.
By definition we have
\begin{equation}
    V_{s}(t_{s}) = 0
\end{equation}

In addition, this corresponds to the situation where evaporation stops exactly when the size of the liquid-vapor interface drops to zero ($2L-4x(t_s) = 0$). Since:
\begin{equation}
    \left. \frac{d}{dt}V_{s}(t)\right\rvert_{t_{s}} = v_e(2L-4x(t_s))
\end{equation}
we have:
\begin{equation}
    \left. \frac{d}{dt}V_{s}(t)\right\rvert_{t_{s}} = 0
\end{equation}

Therefore, searching for the boundary between regions where growth ends before evaporation and the vice versa, means searching for \autoref{equ:solvent} having its minimal value equal to zero. We solve this for $v_e/v_g$ numerically for each $L/h_0$ value.

\subsection{'Intermediate fast' evaporation: evaporation terminates before crystal growth (T $\rightarrow$ E $\rightarrow$ G, formation pathways Nr. 6, 7, 8, red}

ITEG, Nr. 7 and TIEG, Nr. 8: Impingement can take place before or after the crystals touch the film surface. The time for impingement is given by \autoref{equ:ti}  and the time for crystal surface contact by \autoref{equ:Tmeet}, so that the interface is exactly when:
\begin{equation}
    \frac{L/2 - r_{0}}{v_g} = \frac{h_{0} - r_{0}}{v_e + v_g}
\end{equation}

Rearranging: 
\begin{equation}\label{equ:TIIT}
    \frac{v_e}{v_g} = \frac{2h_{0} - L}{L-2r_{0}}
\end{equation}

TEIG, Nr. 6 and TIEG, Nr. 8: Impingement may occur before or after evaporation terminates. The condition for the interface in between is $t_{evap} = t_i$ with $t_i$ given by \autoref{equ:ti}, but this time $t_{evap} = t_{s}$ such that $V_{s}(t_s) = 0$ with, $V_{s}$ defined by \autoref{equ:solvent}. We numerically solve for $v_e/v_g$ to find $t_{s} = t_i$ in order to find the boundary.

\subsection{'Intermediate slow' evaporation: crystal growth terminates before evaporation (T $\rightarrow$ G $\rightarrow$ E, formation pathways Nr. 9, 10, 11), light blue, light and medium yellow}\label{sec:intermediate_slow}

ITGE, Nr 10 and TIGE, Nr. 11: The boundary is equal to \autoref{equ:TIIT}.

TGE, Nr. 9: The crystals do not impinge each other, but the crystals still grow when the crystals touch the vapor (light blue).

After $t_T$ the surface of the crystals that is still surrounded by liquid and where growth further proceed is reduced to $4s(t)$:
\begin{equation}\label{equ:st}
    s(t) = r(t)\text{arcsin}\left(\frac{h(t)}{r(t)}\right)  = (r_0 + v_gt)\text{arcsin}\left(\frac{h_{0} - v_e\cdot t}{r(t)}\right) 
\end{equation}

Considering the 4 growing regions, the total crystalline volume $V_{crystal}(t)$ after $t_T$ reads (as long neither growth nor evaporation terminate):
\begin{equation}\label{equ:Vct}
    V_{crystal}(t) = V_{T} + \int_{t_{T}}^{t} dt\, (4\cdot s(t)\cdot v_g)
\end{equation}

At the end of crystal growth we have:
\begin{equation}
    V_{crystal}(t_{max}) = V_{c,tot}
\end{equation}

Now, growth terminates before the crystals can impinge if:
\begin{equation}
    t_{maxgrowth} < \frac{1}{v_g}\left(\frac{L}{2}-r_{0}\right)
\end{equation}

This means that the limiting condition for end of growth without impingement (film with pinholes) reads:
\begin{equation}\label{equ:Vc}
    V_{crystal}\left(t = \frac{1}{v_g}\left(\frac{L}{2}-r_{0}\right)\right) = V_{sol,tot}
\end{equation}

Using \autoref{equ:Vtot} , \autoref{equ:Tmeet}, \autoref{equ:VT}, \autoref{equ:st},  \autoref{equ:Vct}, \autoref{equ:Vc} this leads to:
\begin{equation}
    \pi (r_{0} + (h_{0}-r_{0})/(v_g+v_e)\cdot v_g)^2 + \int_{(h_{0}-r_{0})/(v_g+v_e)}^{t} dt\, (4\cdot r(t)\text{arcsin}\left(\frac{h_{0} - v_e\cdot t}{r(t)}\right) \cdot v_g) = 2L(1 - \phi_{0})h_0
\end{equation}

For all other parameters fixed, we numerically solve the equation above for $v_e/v_g$. This defines, for each $L/h_0$ value the upper limit on $v_e/v_g$ for a film with pinholes.

\subsection{Roughness}

To calculate the roughness for the complete surface is too complicated. To get an approximation for it we evaluate the highest point of the dry film and compare it to the average film thickness.
\begin{equation}
    R/h_0 = \frac{h_{max}-h_{f}}{h_0}
\end{equation}
For the formation pathways 1-3 the roughness is zero. For the formation pathways 4 and 5 we take the calculated value of $h_{max}$ (compare \autoref{sec:growth}) for the equation above. For the remaining pathways the maximal height of the film is defined by the time the crystals get in contact to the vapor phase. In these cases $h_{max} = r_T$ (comapre \autoref{sec:T}).

\subsection{Uncovered substrate}

There can only be uncovered substrate if no impingement occurs. That leaves formation pathway 4 and 9. For the formation pathway 4 the amount of uncovered substrate can be calculated by 
\begin{equation}
    S/L = \frac{L-2\cdot r(t_{growth})}{L}
\end{equation}
$t_{growth}$ being the crystal radius when the crystal growth terminates, defined in \autoref{sec:growth}.

For formation pathway 9 we can calculate $r(t_{max})$ and the amount of uncovered substrate as above (compare \autoref{sec:intermediate_slow}).

\newpage

\section{Boundary representations with roughness and the amount of uncovered substrate.}

\begin{figure}[h!]
    \centering
    \includegraphics[width = 0.8\textwidth]{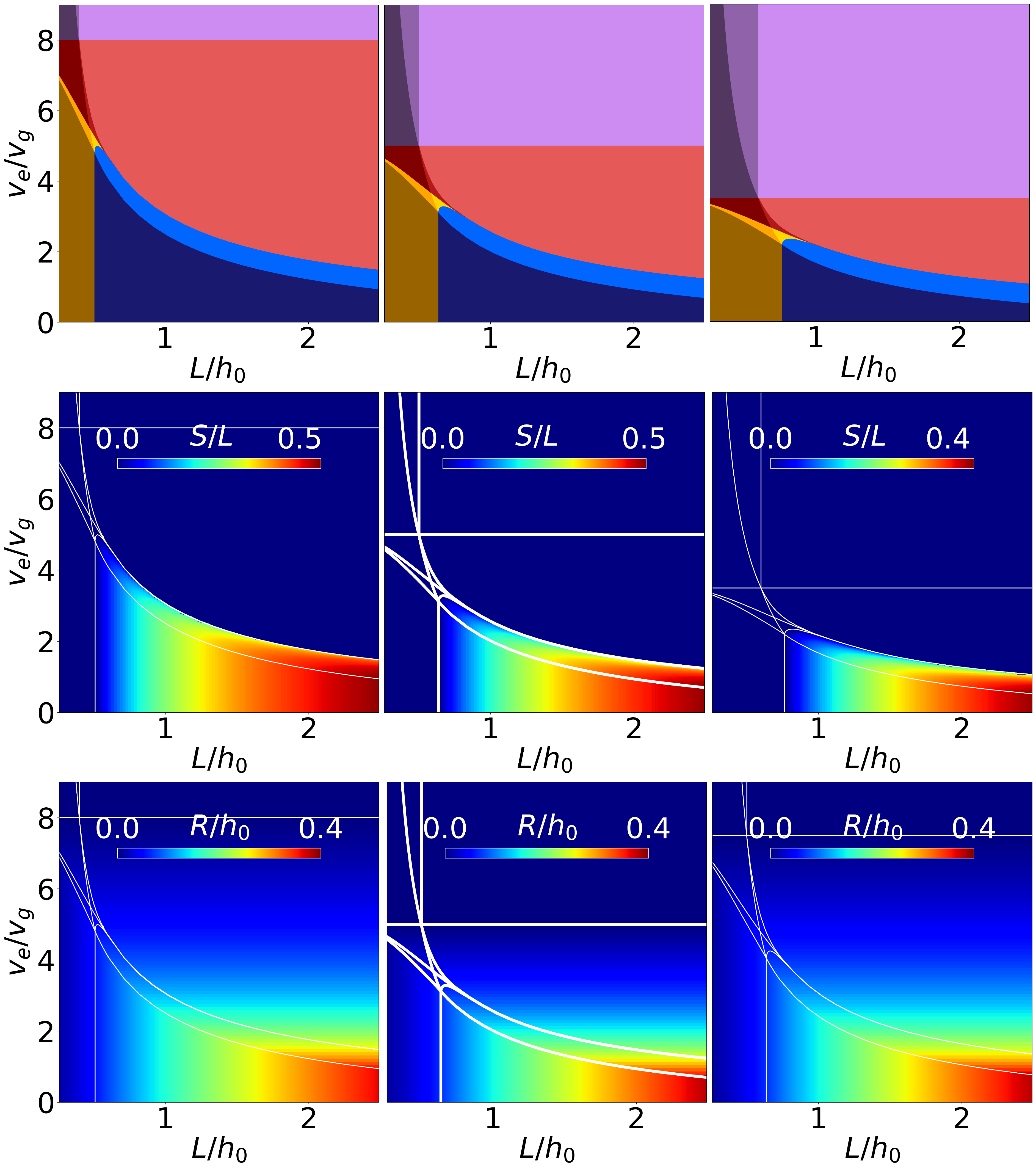}
    \caption{Initial volume fraction ($\phi_{0}$) variation. $\phi_{0}$ increases from left to right ($\phi_{0}$ = 0.2, 0.25, 0.3). The initial ratio of crystal size to film height is $r_0/h_0 = 0.1$. The first row are the boundary representations from the main text. The second and third rows are the corresponding amount of uncovered substrate and roughness plots.}
    \label{fig:phi}
\end{figure}

\begin{figure}[!p]
    \centering
    \includegraphics[width = 0.8\textwidth]{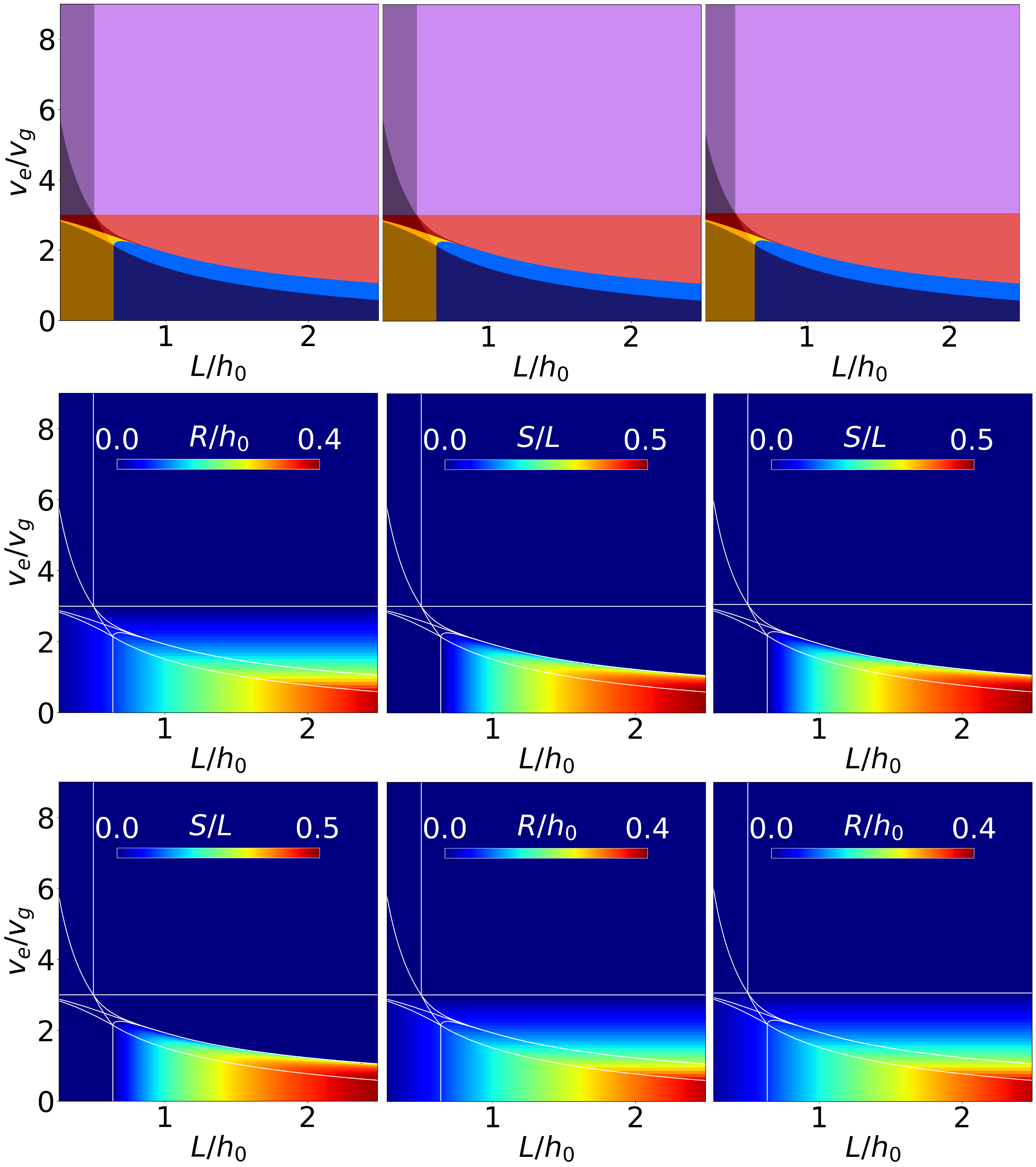}
    \caption{Initial crystal size variation. The ratio $r_{0}/h_{0}$ increases from left to right ($r_{0}/h_{0}$ = $5\cdot10^{-5}$, $5\cdot10^{-4}$, $5\cdot10^{-3}$). The initial volume fraction is $\phi_0 = 0.25$. The first row is the boundary representations from the first row of figure 5 in the main text. The second and third row are the corresponding uncovered substrate and roughness plots.}
    \label{fig:r0small}
\end{figure}

\begin{figure}[h!]
    \centering
    \includegraphics[width = 0.8\textwidth]{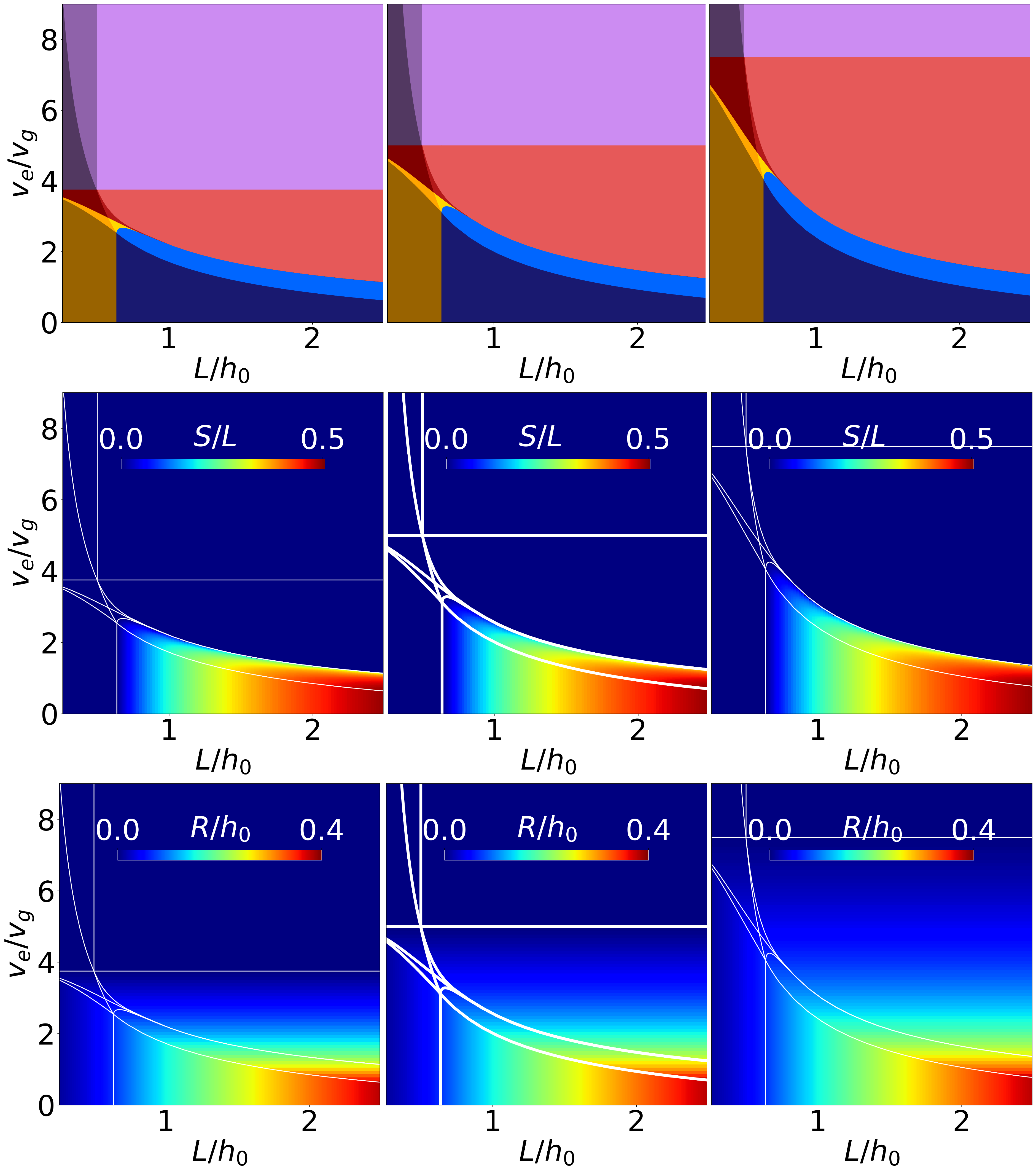}
    \caption{Initial crystal size variation. The ratio $r_{0}/h_{0}$ increases from left to right ($r_{0}/h_{0}$ = 0.05, 0.1, 0.15). The initial volume fraction is $\phi_0 = 0.25$. The first row is the boundary representations from the second row of figure 5 in the main text. The second and third row are the corresponding uncovered substrate and roughness plots.}
    \label{fig:r0large}
\end{figure}

\newpage

\section{Changing the initial film height}

In this section, we discuss the change of the initial film height. In the model, the initial film height is used as a scaling parameter for the x-axis. Additionally, the ratio of initial crystal size to initial film height defines the shape of the boundary representation (compare section 5, main text). For very small ratios of $r_0/h_0$, the effect of changing the ratio is barely visible (compare \autoref{fig:r0small}). Hence, changing only the initial film height will lead to a horizontal displacement in the boundary representation. 

For larger values of $r_0/h_0$, the picture is more complicated. For example: Increasing $h_0$ by a factor of two is equivalent to dividing the initial value of $L/h_0$ by two and could be equivalent to changing the central boundary representation in \autoref{fig:r0large} to the left one. So increasing the film height may (depending on the starting point) move the formation pathway from a film with pinholes to one without (by the horizontal displacement) or it may turn a rough film to a fully smooth one (by the vertical displacement of the interfaces due to the changed boundary representation).

From a practical perspective, a higher initial film height leads to a higher final film height\cite{gumpert_predicting_2023}, which may increase light absorption and therefore enhance efficiency\cite{ouslimane_impact_2021}. The film height can easily be changed for seeded growth as well as heterogeneous nucleation on the substrate. Surprisingly, the necessary evaporation rate to obtain a smooth film (purple region) decreases for an increasing film thickness.

\section{Simulation model}

In this section the equations describing the evolution in the simulation are described.

\subsection{Gibbs Free Energy}

The simulation setup is a simplified version of \cite{ronsin_phase-field_2021}. The system is modeled with three components: the crystallizing material (the solute, $\varphi_1$), the evaporating material (solvent, $\varphi_2$), and the buffer material (air, $\varphi_3$) replacing the solvent. Additionally, the system contains two order parameters to define, whether it is in a fluid state (where both order parameters are zero), in a crystalline state (where the crystalline order parameter $\phi_c = 1$), or in a vapor state (where the vapor order parameter $\phi_{air} = 1$). Finally, there is a labeling field $\theta$ to handle the polycristallinity of the film.

The energy of the system is integrated over the whole system $V$ and is given by the Gibbs free energy $G$
\begin{equation}
    G = \int_V \left( \Delta G^{local} + \Delta G^{interf} \right)dV
\end{equation}
that can be split into a local contribution $\Delta G^{local}$ and a part, that accounts for the surface tensions arising from the interfaces in the system $ \Delta G^{interf}$.

The local contribution can be split into
\begin{equation}
    \Delta G^{local} = \left(1 - \phi_{air}^2(3-2\phi_{air}) \right)\Delta G^{condensed} + \Delta G^{FH} +\phi_{air}^2(3-2\phi_{air})\Delta G^{air} + \Delta G^{numerical}
\end{equation}
where the condensed part $\Delta G^{condensed}$ accounts for the energy difference between the liquid and the solid state, $\Delta G^{FH}$ accounts for the entropic and intramolecular interactions, modeled by the Florry Huggins theory. $\Delta G^{air}$ models the vapor part of the system and $\Delta G^{numerical}$ adds some contributions to enhance numerical stability. $\Delta G^{condensed}$ can be written as:
\begin{equation}
    \Delta G^{condensed} = \rho\varphi_1^2 \left( \phi_c^2(1-\phi_c)^2W + \phi_c^2(3-2\phi_c)\Delta G^{cryst} \right)
\end{equation}
where $\rho$ is the density of the materials, $W$ defines the energy barrier between fluid and solid state, and $\Delta G^{cryst} = L\left( T/T_m - 1 \right)$, with $L$ being the enthalpy of fusion, $T$ the temperature and $T_m$ the melting temperature of the perovskite.

The intramolecular interactions between the species $\Delta G^{FH}$ can be written as
\begin{equation}
    \Delta G^{FH} = \frac{RT}{\nu_0}\left( \sum_{i=1}^3 \varphi_i ln(\varphi_i) + \sum_{i=1}^3\sum_{j>i}^3 \varphi_i\varphi_j\chi_{ij,LL} + \sum_{j=2}^3\phi_c^2\varphi_1\varphi_j\chi_{1j,SL} \right)
\end{equation}
where $R$ is the gas constant, $\nu_0$ is the volume of the smallest lattice site, as defined in the Florry Huggins theory, $\chi_{ij,LL}$ accounts for the interactions between material $\varphi_i$ and material $\varphi_j$ and $\chi_{1j,SL}$ correspondingly between the solid part of material $\varphi_1$ and material $\varphi_j$. 

The energy of the air is modeled as an ideal gas:
\begin{equation}
    \Delta G^{air} = \frac{RT}{\nu_0}\sum_{i=1}^3\varphi_iln\left( \frac{\varphi_i}{\varphi_{sat,i}} \right)
\end{equation}
where $\varphi_{sat,i} = P_{sat,i}/P_0$ with $P_{sat,i}$ being the vapor pressure of material $\varphi_i$ and $P_0$ a reference pressure. The numerical contribution can be written as:
\begin{equation}
    \Delta G^{numerical} = E_0\frac{d_{sv}}{f(\varphi_i\phi_c,d_{sv},c_{sv},w_{sv})} + \sum_{i=1}^3\frac{\beta}{\varphi_i}
\end{equation}
where the first summand prevents the crystalline and the vapor phase from penetrating each other, with $E_0$ defining the strength of the penalty and $f$ being an interpolating function and $d_{sv},c_{sv},w_{sv}$ being its amplitude, center, and width as defined below. The second summand helps to increase the possible step size of the simulation by preventing the volume fractions from becoming too small, where $\beta$ defines the strength of this penalty and it is chosen small enough to have negligible impact on the physical properties. The interpolating function is defined as
\begin{equation}
    \text{log}f(x,d,c,w) = \frac12 \text{log}(d)(1+\text{tanh}(w(x-c)) )
\end{equation}

The surface tension arises from the interfacial contribution
\begin{equation}
    \Delta G^{interf} = \sum_{i=1}^3\frac{\kappa_i}{2}\left(\nabla\varphi_i\right)^2 + \frac{\epsilon_{air}^2}{2}(\nabla \phi_{air})^2 + \frac{\epsilon_c^2}{2}\left( \nabla \phi_c \right)^2 + p(\phi_c)\frac{\pi\epsilon_g}{2} \delta\left(\nabla\theta\right)
\end{equation}
where $\kappa_i$ defines the surface tension arising from volume fraction variations of material $\varphi_i$, $\epsilon_{air}$ defining the contribution of the interface between the vapor and non-vapor phase, $\epsilon_{c}$ defining the contribution arising from the crystalline - noncrystalline interface, and $\epsilon_g$ accounting for the energy contribution of crystal boundaries.

\subsection{Allen Cahn and Cahn Hilliard Equation}

The evolution of the volume fractions is modeled by the Cahn-Hilliard equation
\begin{equation}
    \frac{\partial\varphi_i}{\partial t} = \frac{\nu_0}{RT} \nabla \left[ \sum_{j=1}^2\Lambda_{ij} \nabla (\mu_j - \mu_3)  \right]
\end{equation}
where $\Lambda_{ij}$ are the onsager mobilities and $\nabla (\mu_j - \mu_3)$ is the difference in chemical potential. The evolution of the order parameters is given by the Allen Cahn equation
\begin{equation}
    \frac{\partial \phi_i}{\partial t} = -\frac{\nu_0}{RT}M_i\frac{\delta\Delta G}{\delta \phi_i}
\end{equation}
where $i$ stands for $air$ or $c$ for the vapor or crystalline phase respectively, and $M_i$ is the respective Allen Cahn mobility. The evaporation of the solvent is modeled as an outflux $j$ at the top of the simulation box $z=z_{max}$
\begin{equation}
    j^{z=z_{max}}=\alpha \sqrt{\frac{\nu_0}{2\pi RT\rho}}P_0(\varphi_2^{vap} - \varphi^{\infty})
\end{equation}
where $\alpha$ is the evaporation condensation coefficient and $\varphi_2^{vap}$ is the vapor pressure of the solvent and $\varphi^{\infty}$ the vapor pressure of the environment. It was ensured that this outflux reproduces the evaporation rates measured in the experiment\cite{ronsin_phase-field_2021}.

\newpage

\section{Parameters (Simulation)}

\begin{table}[h!]
\caption{Parameters of the simulations}
    \centering \small
        \begin{tabular}{ |p{3.1cm}| p{5.1cm}| p{3.1cm}|p{3.1cm}| }\hline
        \rowcolor{lightgray} Parameters & Full Name & Value & Unit\\ \hline
         $\alpha$ & \makecell[l]{Evaporation-\\condensation-\\coefficient} & $2.3\cdot 10^{-5}$ & - \\ \hline
         dx, dy & Grid Spacing & 1, 1 & nm \\  \hline
        $\phi_{1,ini}$& \makecell[l]{Initial precursor\\ concentration }& 0.20, 0.25 or 0.3 & - \\ \hline
        nx & Grid Size in horizontal direction & 160, 192, 224, 256, 288, 320, 352, 384, 416  & - \\ \hline
        ny & Grid Size in vertical direction & 256  & - \\ \hline
        T & Temperature & 300 & K \\ \hline
        $\rho$ & Density & 1000 & $kg/m^3$ \\ \hline
        m & Molar Mass & 0.1, 0.1, 0.03 & $kg/mol$ \\ \hline
        $\nu_0$& \makecell[l]{Molar Volume \\of the Florry\\ Huggins Lattice Site} & $3\cdot 10^{-5}$ & $m^3/mol$ \\ \hline
        $\chi_{12,LL},\chi_{13,LL},\chi_{23,LL}$& Liquid - liquid interaction parameter & 0.57, 1, 0 & - \\ \hline
        $\chi_{12,SL},\chi_{13,SL}$&Liquid - solid interaction parameter & 0.15, 0.5 & - \\ \hline
        $T_m$ & Melting Temperature & 600 & $K$ \\ \hline
        $L_{fus}$& Heat of Fusion & 75789 & $J/kg$ \\ \hline
        W & Energy barrier upon crystallization & 142105 & $J/kg$ \\ \hline
        $P_0$& Reference Pressure & $10^5$ & $Pa$ \\ \hline
        $P_{sat,1},P_{sat,2},P_{sat,3}$& Vapor Pressure & $10^2$, $1.5\cdot 10^4$, $10^8$ & $Pa$ \\ \hline
        $P_i^\infty$& Partial Vapor Pressure in the Environment & 0 & $Pa$ \\ \hline
        $E_0$& Solid-Vapor interaction energy & $5\cdot 10^9$ & $J/M^3$\\ \hline
        $\beta$& Numerical Free Energy Coefficient & $10^{-5}$ & $J/M^3$ \\ \hline
        $\kappa$ & Surface Tension Parameters for Volume Fraction Gradients & $6\cdot 10^{-9}$ (all) & $J/m$ \\ \hline
        $\epsilon_c,\epsilon_{vap}$& Surface Tension Parameters for Order Parameter Gradients & $3\cdot10^{-5}$, $10^{-4}$& $\left(J/m\right)^{0.5}$\\ \hline
        $D_{s,i}^{\Phi_j\rightarrow 1}$& Self-Diffusion Coefficients in pure materials & $10^{-9}$ (all) & $m^2/s$ \\ \hline
        $M_c$& Allen Cahn mobility coefficient for the crystalline phase & 0.5, 0.75, 1, 1.5, 2, 2.5, 3 & $s^{-1}$ \\ \hline
        $M_v$& Allen Cahn mobility coefficient for the vapor phase & $10^6$ & $s^{-1}$ \\ \hline
        $D_1^{vap},D_2^{vap},D_3^{vap}$& Diffusion Coefficients in the Vapor Phase & $10^{-18}$, $10^{-10}$, $10^{-10}$ & $m^2/s$\\ \hline
        $d_{sl},c_{sl},w_{sl}$& Amplitude, center and with of the penalty function for the diffusion coefficients upon liquid solid transition & $10^{-9}$, 0.7, 10 & -\\ \hline
        $d_{sv},c_{sv},w_{sv}$& Amplitude, center and with of the penalty function for the Allen Cahn mobility and the solid- vapor interaction energy & $10^{-9}$, 0.3, 15 & -\\ \hline
        $r_{ini}$& Radius of the initially placed crystals & 10, 20, 30 & nm\\ \hline
    \end{tabular}
    
    \label{tab:Param}
\end{table}

\normalsize

\newpage

\section{Crystal Growth Rate in the Simulation}\label{sec:CrGrRate}

\begin{figure}[h!]
    \centering
    \includegraphics[width = 0.7\textwidth]{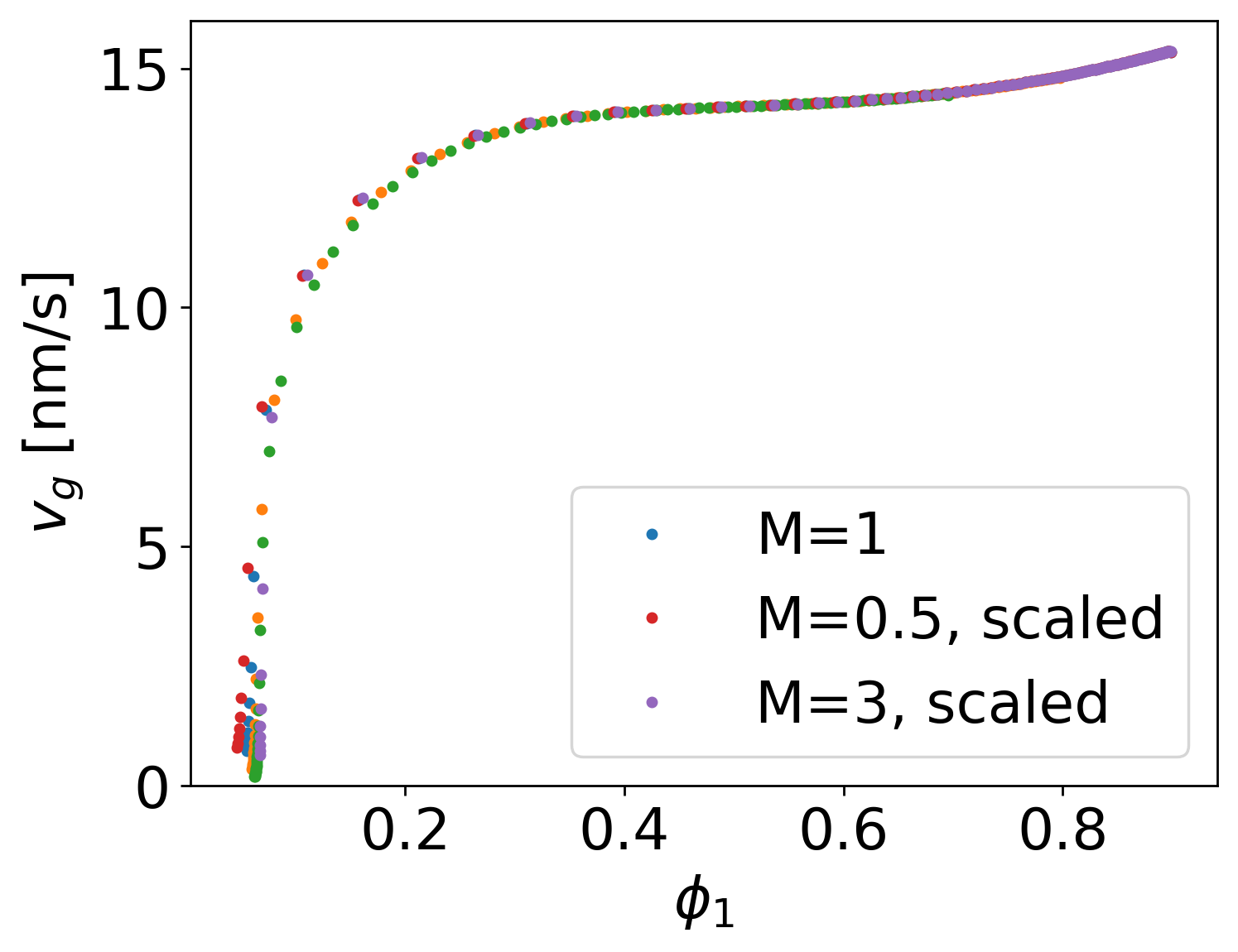}
    \caption{Grow Rates depending on the solute volume fraction $\Phi_1$ evaluated from a one-dimensional simulation with a singular crystal placed. Three different Allen-Cahn mobilities were tested to ensure consistency. The curves for M = 0.5 and M = 3 are scaled with 2 and 1/3.}
    \label{fig:GrRate}
\end{figure}

\section{Verification of the theoretical model against PF Simulations}

The validity of the model is cross checked against a PF simulation\cite{ronsin_phase-field_2021}. The simulation takes into account the three components of solute, solvent and vapor, and captures evaporation, crystal growth and diffusion. The initial arrangement is the same as that used in the theoretical model. The main differences between the two approaches are: (1) the interfaces between the phases are sharp in the theoretical model but diffuse in the PF approach; (2) surface tensions play a role in the PF simulation; and (3) in the PF simulation the crystal growth rate and the solvent evaporation rate weakly depend on the surrounding volume fraction. Additionally, the crystalline phase consists a residue amount of solvent, and crystal growth stops at a volume fraction of a few percent solute (at the thermodynamic equilibrium).

The visual representation is shown in \autoref{fig:comparison}. Between simulation runs/minimal model evaluations, we vary the ratio of evaporation to growth rate $v_e/v_g$ and the distance between the crystals $L$. We scale these distance with the initial film height $h_{ini}$ to obtain a dimensionless quantity. In \autoref{fig:comparison}, simulations are represented by circles, and the background color indicates the morphology classification scheme. First, all the formation pathways can be recovered in the simulation. Additionally, there is a good agreement between theory and simulation, although slight deviations are observed. The reasons for these deviations are: 

\begin{figure}[h]
\centering
  \includegraphics[width = 0.45\textwidth]{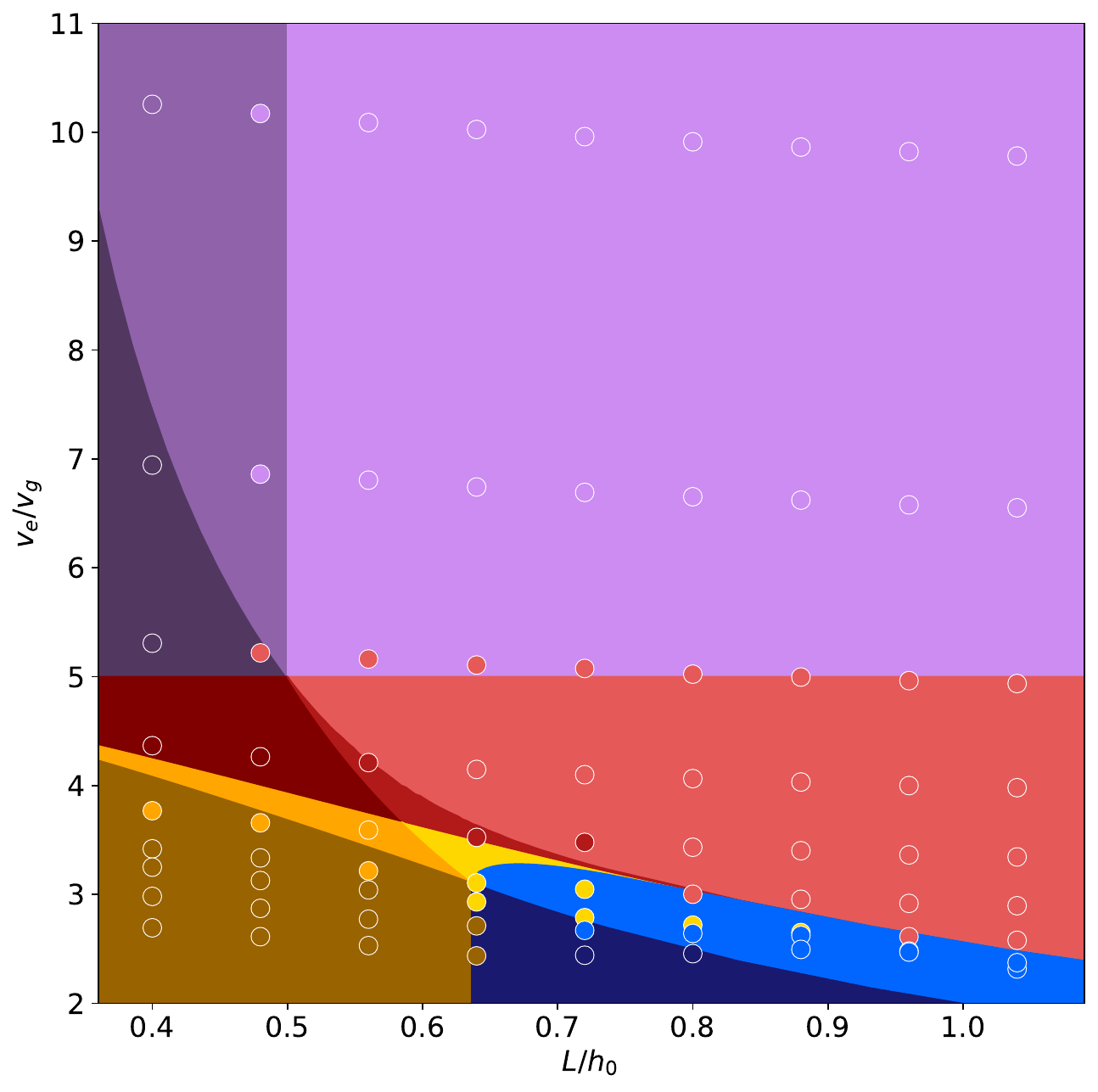}
  \caption{Comparison between the model and the results of the Phase Field simulations. The initial volume fraction of crystalline material is $\phi_0 = 0.25$, and the initial ratio of crystal size to film height is $r_0/h_0 = 0.1$ (same as in Figure 2/3 in the main text). The simulations are represented as dots. There is an overall good agreement.}
  \label{fig:comparison}
\end{figure}

First, in the simulations, the growth rate is not constant for all volume fractions (see \autoref{sec:CrGrRate}). To partially compensate for this, the growth rate of the crystals is averaged from the initial to the dry state. This is achieved by tracking the composition in the wet film and averaging the interface velocities obtained from the 1D crystal growth simulations in \autoref{sec:CrGrRate}. Although this simple correction improves the agreement, it is not sufficient to fully compensate for all the differences with the minimal model. This leads to a slight inaccuracy in the value of $v_e/v_g$.

Second, in the Phase Field simulation, a surface tension between the different materials and phases must be included. This leads to a minimal possible gap distance larger than zero between the crystals. Consequently, the boundary for low $v_e/v_g$ is shifted to larger $L/h_0$ values compared to the theoretical model. Additionally, the Phase Field simulation contains diffuse interfaces, also preventing very small gaps between the crystals (region of light blue).

Further effects that contribute to small deviation to the theory include: the volume fraction of every material at any point in the simulation cannot be exactly zero, and the evaporation rate is not perfectly constant in the simulation\cite{ronsin_phase-field_2021}.


\bibliographystyle{unsrt}
\bibliography{FP-paper}

\end{document}